\documentclass[useAMS,usenatbib]{mn2e} 
\usepackage{epsfig}
\def\equ#1{eq.~(\ref{eq:#1})}
\def\Equ#1{Eq.~(\ref{eq:#1})}
\def\se#1{\S\ref{sec:#1}}
\def\fig#1{Fig.~\ref{fig:#1}}
\def\Fig#1{Fig.~\ref{fig:#1}}
\def\be{\begin{equation}}
\def\ee{\end{equation}}

\def\etal{{\it et al.\ }}
\def\msun{M_{\odot}}
\def\solm{M_{\odot}}

\def\ifm#1{\relax\ifmmode#1\else$\mathsurround=0pt #1$\fi}
\def\kms{\ifmmode\,{\rm km}\,{\rm s}^{-1}\else km$\,$s$^{-1}$\fi}

\def\ltsima{$\; \buildrel < \over \sim \;$}
\def\lsim{\lower.5ex\hbox{\ltsima}}
\def\gtsima{$\; \buildrel > \over \sim \;$}
\def\gsim{\lower.5ex\hbox{\gtsima}}
\def\prop{\propto}

\def\gameff{\gamma_{\rm eff}}
\def\gamthr{\gamma_{\rm crit}}

\def\Rd{R_{\rm disc}}
\def\rs{r_{\rm s}}
\def\us{u_{\rm s}}
\def\cs{c_{\rm s}}
\def\del{\delta}
\def\bu{{\bf u}}
\def\bnabla{{\bf \nabla}}
\def\omm{\Omega_{\rm m}}
\def\oml{\Omega_{\Lambda}}
\def\Dvir{\Delta_{\rm v}}
\def\lcdm{$\Lambda$CDM }

\def\fb{f_{\rm b}}
\def\lya{$L_\alpha$ }

\def\zv{z_{\rm v}}
\def\rv{r_{\rm v}}
\def\vv{v_{\rm v}}
\def\tv{t_{\rm v}}
\def\ai{a_{\rm i}}
\def\ri{r_{\rm i}}
\def\ti{t_{\rm i}}
\def\zi{z_{\rm i}}
\def\ci{C_{\rm i}}
\def\eti{\eta_{\rm i}}
\def\deli{\delta_{\rm i}}
\def\rhuu{\rho_{\rm u}}
\def\rhui{\rho_{\rm ui}}
\def\rhu0{\rho_{\rm u0}}
\def\r1i{r_{\rm 1}}
\def\av{a_{\rm v}}
\def\delv{\delta_{\rm v}}

\def\ap{a_{\rm p}}
\def\ag{a_{\rm g}}
\def\md{M}
\def\mg{m}
\def\mt{{\cal M}}

\title[Virial shock in haloes]
{Virial shocks in galactic haloes?}   
\author[Y. Birnboim \& A. Dekel]
{Yuval Birnboim \& Avishai Dekel\\ 
Racah Institute of Physics, The Hebrew University, Jerusalem Israel\\
yuval@frodo.fiz.huji.ac.il; dekel@phys.huji.ac.il}

\begin{document}

\pagerange{\pageref{firstpage}--\pageref{lastpage}} \pubyear{2002}
\maketitle
\label{firstpage}

\begin{abstract}
We investigate the conditions for the existence of an expanding virial 
shock in the gas falling within a spherical dark-matter halo.
The shock relies on pressure support by the shock-heated gas behind it.
When the radiative cooling is efficient compared to the infall rate
the post-shock gas becomes unstable; it collapses inwards
and cannot support the shock.
We find for a monoatomic gas that the shock is stable
when the post-shock pressure and density obey 
$\gameff\equiv (d\ln P/dt)/(d\ln \rho/dt)>10/7$. 
When expressed in terms of the pre-shock gas properties at radius $r$ 
it reads $\rho r \Lambda(T) /u^3 <.0126$, 
where $\rho$ is the gas density, $u$ is the infall
velocity and $\Lambda(T)$ is the cooling function, with the
post-shock temperature $T\prop u^2$.
This result is confirmed by hydrodynamical simulations, using an accurate 
spheri-symmetric Lagrangian code.
When the stability analysis is applied in cosmology,
we find that a virial shock does not develop in most haloes that form
before $z\sim 2$, and it never forms in haloes less massive than a few 
$10^{11}\solm$.  
In such haloes, the infalling gas is not heated to the 
virial temperature until it hits the disc, 
thus avoiding the cooling-dominated quasi-static contraction phase.
The direct collapse of the cold gas into the disc
should have nontrivial effects on the star-formation rate and on outflows.
The soft X-ray produced by the shock-heated gas in the disc is expected
to ionize the dense disc environment, and the subsequent recombination 
would result in a high flux of \lya emission. 
This may explain both the puzzling low flux of soft X-ray background and the
\lya emitters observed at high redshift.
\end{abstract}


\section{Introduction}
\label{sec:intro}

The standard lore in the idealized picture of galaxy formation by spherical 
infall of gas inside dark-matter haloes is that the gas is first heated to 
the halo virial temperature behind an expanding virial shock. 
It is then supported by pressure in a quasi-static equilibrium  
while it is cooling radiatively and is slowly contracting
to a disc where it can eventually 
form stars.  The cooling process thus
determines important galaxy properties such as the star-formation rate and
the metal enrichment, so it is necessarily an important ingredient in the
galaxy formation process. 

However, it is not at all clear that a stable shock can persist in the 
halo gas away from the disc under the conditions valid in many galactic haloes.
In the absence of a virial shock, the gas is not heated to the virial 
temperature until it falls all the way to the disc, where the collapse 
stops and the gas is heated in a thin layer. 
This may alter some of the assumed processes of disc formation 
and in particular the star formation rate in it. 
It may work against blowout by supernova-driven winds in dwarf galaxies. 
The result of heating near the disc instead of at the virial radius may result
in weakening the soft x-ray emission from such haloes and producing a 
high flux of \lya instead.
In this paper we evaluate the conditions for the existence of a virial 
shock in galactic haloes.

Initial density perturbations are assumed to grow by gravitational instability,
reach maximum expansion, and collapse into virial equilibrium at roughly half
the maximum-expansion radius.
During the initial phase, and roughly until shells start crossing each other
near the virial radius, the gas pressure is negligible compared to the 
gravitational force, so the shells of gas and dark matter move in a 
similar manner. Once interior to the virial radius, 
where shells tend to cross and the gas density becomes high enough, 
the gas pressure becomes an important player in the dynamics.
Its hydrodynamic properties allow transfer of bulk kinetic energy into 
internal energy and the pressure prevents gas element from passing 
through other gas elements and from being compressed without limit.
This makes the infall velocity vanish at the centre.
Since in the cold infalling gas the typical velocity is higher than 
the speed of sound, the information about this inner boundary
condition cannot propagate outwards in time, and these supersonic
conditions create a shock. After the gas crosses the shock, it is heated up,
the speed of sound increases, and the flow becomes subsonic.

The shock transfers the kinetic energy that has been built during the 
collapse into internal gas energy just behind the shock. 
A stable spherical shock would slowly propagate outwards through the 
infalling gas, leaving behind it hot, high-entropy gas that is almost at rest. 
The temperature of the post-shock gas roughly equals the virial temperature. 
The persistence of the shock depends on sufficient pressure by the post-shock
gas, which supports it against being swept inwards due to the gravitational
pull together with the infalling matter. Radiative gas cooling  
makes the gas lose entropy and pressure, which weakens the pressure support 
behind the shock front. Our approach here is to evaluate the existence
of a virial shock by analyzing the gravitational stability of the supporting
gas behind the shock in the presence of significant cooling. 

In \se{analytical} we first summarize the standard analysis of an
adiabatic shock and then generalize the gravitational
stability criterion to the case where cooling is important.
In \se{code} we describe our spherical hydrodynamic Lagrangian code,
which includes gravitating dark-matter and gas shells, artificial
viscosity, radiative cooling and centrifugal forces. We test the code 
in this section and in Appendix \ref{app:code}.
In \se{simu} we apply the numerical code to simulations which demonstrate
the shock formation and test the validity of the analytical model. 
In \se{cosmology} we apply the shock stability criterion to realistic
haloes forming in cosmological conditions.
In \se{conc} we summarize our results and discuss potential astrophysical
implications.

\section{Shock stability analysis}
\label{sec:analytical}

Our goal here is to derive a criterion for the existence of a virial shock
in terms of the properties of the infalling gas just in front of
the shock front. It is based on a gravitational stability analysis of 
the post-shock gas. 
We first remind ourselves of the standard stability analysis in the 
simple adiabatic case, and then derive a more general criterion
for stability in the radiative case, under certain assumptions 
and using a perturbation analysis.

\subsection{The standard adiabatic case}
\label{sec:nocool}

Throughout this paper, we treat the baryons as an ideal monoatomic gas. 
Their {\it equation of state} could therefore be written as
\be
\label{eq:ideal_eos}
P=(\gamma-1)\, e\, \rho \, ,
\ee
where $P$ is the pressure, $e$ is the specific internal energy,
$\rho$ is the density of the gas and $\gamma$ is the adiabatic index. 
Along an isentrope (an adiabatic process of constant entropy) 
the pressure and density are related via $P \propto \rho^\gamma$,
so the adiabatic index is defined by
\be
\label{eq:gamma}
\gamma = \left({\partial \ln P \over \partial \ln \rho}\right) _{\rm s} \,
. 
\ee
For a monoatomic gas $\gamma = 5/3$.\footnote{As
the temperature exceeds the binding energy of the hydrogen and helium
atoms,
electrons become detached from
the nuclei and $\gamma$ becomes smaller.  Once the gas becomes
fully ionized, the original value of $5/3$ is restored, but with a
different effective density. This should have only a marginal effect on
our results, and is ignored in this paper.}

The virial shock is assumed to be a spherical accretion shock which
propagates outwards slowly while infalling gas crosses it inwards.

The kinetic energy of the infalling gas is transformed at the shock front
into thermal energy ---  the post-shock gas is thus heated to a
temperature 
close to the virial temperature of the system of dark-matter halo and
gas,
$V_{\rm infall}^2\approx k_{\rm B} T_{\rm vir}$. 
Because the original temperature of the infalling gas is negligible
compared to the virial temperature, the system obeys the strong-shock
limit.
When we denote the pre-shock and post-shock quantities by subscripts 0 and
1 respectively, the jump conditions across the shock are in this case
\citep{zelrai}:
\be
\rho_0=\frac{\gamma -1}{\gamma +1} \, \rho_1 \,, 
\label{eq:j1}
\ee
\be
(u_0-\us)=\frac{\gamma +1}{\gamma -1} \, (u_1-\us) \, , 
\label{eq:j2}
\ee
\be
P_1=\frac{2\rho_0 u_0^2}{\gamma+1} \, ,
\label{eq:j3}
\ee
\be
T_1 =\frac{\mu}{k_bN_a}\frac{P_1}{\rho_1}
=\frac{\mu}{k_bN_a} \frac{2\gamma-1}{(\gamma+1)^2} u_0^2 \, ,
\label{eq:j4}
\ee
where $u$ stands for radial velocity, $\us$ is the shock velocity,
$N_a$ is Avogadro's number, ${N_a}/{\mu}$ is the average number of
molecules per unit mass, and $k_B$ is Boltzman's constant.

According to standard shock theory, the post-shock gas is always sub-sonic
(in the frame of reference of the moving shock)
because of the increase of the sound velocity behind the shock. 
This gas is thus capable of providing
the necessary pressure to support the shock against the gravitational
pull
inwards applied by the self-gravity of the gas and the dark-matter halo
as well as the pressure applied by the infalling matter at the shock
front. 
 
The criterion for gravitational stability
of this post-shock gas in the adiabatic case
is the standard Jeans stability criterion:  
$\gamma > 4/3$ \citep[e.g., ][Chapter 8]{cox:80}

If the post-shock gas is gravitationally unstable, it falls into the 
galaxy centre on a dynamical timescale and can no longer support the shock.
As a result, the shock weakens and it is swept inwards.

The Jeans criterion can be qualitatively understood in terms of the 
following heuristic derivation. For a shell of radius $r$, we compare 
the gravitational pull inwards, $\ag=GM/r^2$ (where $M$ is the
mass interior to $r$),  
to the pressure pushing outwards, $\ap=\rho^{-1}\nabla P$.
We assume an isentrope, $P \prop \rho^\gamma$. We also assume homology,
such 
that the local density scales like the mean density in the sphere
interior
to $r$, $\rho \prop M/r^3$.
Then $\nabla P$ can be replaced by $\sim P/r$ and we obtain
\be
\frac{\ap}{\ag} \prop \rho^{\gamma-4/3} \, .
\label{eq:jeans}
\ee
If $\gamma<4/3$, we have an unstable configuration.
Starting in hydrostatic equilibrium, $\ap/\ag=1$,
a perturbation involving contraction is associated with a larger 
$\rho$, and therefore $\ap/\ag<1$ by \equ{jeans}, 
implying that the pressure cannot prevent collapse.
If $\gamma>4/3$, the pressure force 
increases until it balances the increased gravitational pull.
We note that even this simple derivation of the Jeans criterion
had to assume homology --- an assumption that we will have to adopt
also in our analysis of the radiative case below.
 
\subsection{Shock stability under radiative cooling}
\label{sec:cool}

We wish to replace the adiabatic Jeans criterion by a more general
stability condition that will be valid also in the radiative case.
This criterion must depend on the cooling rate and should therefore
be naturally expressed in terms of time derivatives.
We generalize the adiabatic $\gamma$ of \equ{gamma}
by an effective $\gamma$ following a comoving volume element
along its Lagrangian path:
\be
\label{eq:def_gam_eff}
\gameff \equiv { d \ln P/dt \over d \ln\rho/dt } \, .
\ee
We expect that the system would be stable when $\gameff$ is larger than 
a certain critical value, the analog of the requirement $\gamma>4/3$ 
in the adiabatic case.

In our Lagrangian analysis all the quantities
($r$, $u$, $\rho$, $P$, etc.) refer to comoving shells; they
are all functions of the gas mass $m$ interior to radius $r$ 
and time $t$. Derivatives with respect to time following a comoving
volume
element will be denoted by an upper dot, and derivatives with respect to
m
will be denoted by a prime.

The effective gamma can be related to its adiabatic analog given the
cooling
rate and other post-shock gas quantities.  The time derivative
of \equ{ideal_eos} yields:
\be
\label{eq:dpdt}
\dot{P} = (\gamma-1) (\dot{e} \rho +e\dot\rho) \, .
\ee
Energy conservation in the presence of radiative losses can be
expressed by
\be
\label{eq:dedt}
\dot e=-P \dot V -q=\frac{P\dot\rho}{\rho^2} -q \, ,
\ee
where $q$ is the radiative cooling rate [to be discussed below, 
e.g., \equ{q_def}] and $V=\rho^{-1}$ is the specific volume.
Substituting $\dot e$ from \equ{dedt} into \equ{dpdt}, and using it in
\equ{def_gam_eff}, we obtain  
\be
\label{eq:fin_gam_eff}
\gameff=\gamma-\frac{\rho}{\dot\rho} \frac{q}{e} \, .
\ee
Note that in the limit $q/e\ll\dot\rho/\rho$ we reproduce the adiabatic
case; the process is nearly adiabatic when the cooling timescale is long 
compared to the contraction timescale. 

We assume that in the region close behind the shock the pattern 
of the velocity field is {\it homologous}. 
By this we mean that at any given time the 
(radial) velocity is proportional to the radius (as in a Hubble flow),
namely
\be
\label{eq:homo}
u/r=u_1/\rs \, ,
\ee
thus providing a boundary condition for the post-shock gas. 
The homology is shown to be a valid approximation in the simulations
discussed below, where the post-shock shell trajectories are roughly
parallel to each other in the $\log r - t$ plane, at any given time
close enough to shock crossing.
The time evolution of the density can then be evaluated via the 
{\it continuity} equation in Lagrangian form for the spheri-symmetric
case,
\be
\label{eq:drodt}
\frac{\dot\rho}{\rho} =-\bnabla \cdot \bu 
=-\frac{1}{r^2}\frac{\partial}{\partial r}(r^2 u)
=-\frac{3 u_1}{\rs} \, ,
\ee
where the last equality results from the assumed homology, \equ{homo}.
The homology thus implies that $\dot \rho /\rho$ at a given $t$
is a constant in $m$ throughout the post-shock region.
\Equ{fin_gam_eff} can then be simplified:
\be
\gameff = \gamma + \frac{\rs}{3u_1} \frac{q}{e} \, .
\label{eq:gameff}
\ee

We start with a hypothetical unperturbed state for the post-shock gas,
where we assume that the net force {\it vanishes}, $\ddot r=0$.
The system adjusts itself to this state on a timescale associated with  
the speed of sound $\cs$, provided that it is much higher than the
infall velocity $u$. This is expected to be the case in the sub-sonic
post-shock medium, where $\cs$ becomes high and $u$ becomes low.
The unperturbed equation of {\it motion} in Lagrangian form is then
\be
\label{eq:dpdm} 
\ddot{r} = -4\pi r^2 P' -\frac{GM}{r^2} = 0 \, ,
\ee
where $M$ is the total mass interior to radius $r$.

We then introduce a perturbation due to a homologous infall velocity
$u$.
Over a short time interval $\del t$, it
introduces a small displacement inwards, $\del r = u\, \del t$.
In order to distinguish between stability and instability we wish
to determine whether the induced acceleration, 
$\ddot{\del r}$, is positive or negative, tending to decrease or increase
the
velocity respectively.
Note that under homology, \equ{homo}, the relative displacement is 
\be 
{\del r}/{r} = {u_1 \,\del t}/{\rs} \, .
\label{eq:rad} 
\ee

Writing the equation of motion, \equ{dpdm},
but for the perturbed quantities $P+\del P$ and $r+\del r$, 
and subtracting the unperturbed \equ{dpdm}, we obtain to first order
\be
\ddot{\del r} = -4\pi r^2 (\del P)' +\frac{4 GM\, \del r}{r^3} \, .
\label{eq:motion_first}
\ee
We next manipulate the right-hand side of \equ{motion_first} to obtain
a simple expression involving $\gameff$.

In the second term we use the homology, \equ{rad},
and then the unperturbed equation of motion, \equ{dpdm}, to obtain
\be
\frac{4 GM\, \del r}{r^3}
=-\frac{16\pi r^2 u_1 \del t}{\rs} P' \, .
\label{eq:term2}
\ee

The manipulation of the first term is somewhat more elaborate.
We use the definition of $\gameff$, \equ{def_gam_eff}, to write
\be
\del P = (\dot \rho/\rho) P \gameff \del t
  = - (3 u_1/\rs) P \gameff \del t \, ,
\ee
where the second equality is due to \equ{drodt}.
Note that the $m$ dependence in this term is only in the product
$P\gameff$.
We now express $\gameff$ in terms of the cooling rate $q$
as in \equ{gameff}, and need to take the derivative $(Pq/e)'$.
We make here the standard assumption that the radiative cooling rate is
proportional to density,
\be
q=\rho\Lambda (T) \, ,
\label{eq:q_def}
\ee
with $\Lambda (T)$ the macroscopic cooling function and
$T$ the post-shock temperature.
The immediate post-shock medium is assumed to be isothermal,
reflecting via the jump conditions an assumed approximate uniformity 
of the pre-shock gas over a short time interval.
Using \equ{ideal_eos} we have $P/e=(\gamma-1)\rho$, and together
with \equ{q_def} it becomes  
\be
Pq/e=(\gamma -1) \Lambda \rho^2 \, .
\label{eq:Pqe}
\ee
In the computation of $(Pq/e)'$, we first replace $\rho'=(d\rho/dP)P'$, 
then use \equ{def_gam_eff} to write $d\rho/dP = \gameff^{-1} \rho/P$,
use \equ{Pqe} backwards to replace $(\gamma -1) \Lambda \rho^2/P$ by
$q/e$,
and finally use \equ{gameff} to obtain
$(Pq/e)' = 3u_1/r_s[-2 \gameff^{-1} (\gamma -\gameff) P']$.
We thus have in the first term of the rhs of \equ{motion_first} 
\be
-(\del P)' 
= \frac{3u_1\del t}{\rs} P' [\gamma - 2\gameff^{-1}(\gamma-\gameff)] \,
.
\label{eq:term1}
\ee

With the right-hand side of \equ{motion_first} given
by \equ{term1} and \equ{term2}, the first-order equation finally
becomes,
\be
\ddot{\del r} = \frac{12\pi r^2 u_1\del t P'}{\rs}  
\left[\gamma - 2\gameff^{-1}(\gamma-\gameff) -\frac{4}{3}\right]  \, .
\ee
Since $u_1$ and $P'$ are both always negative,  
the desired sign of $\ddot{\delta r}$ is determined by the sign of
the expression inside the square brackets.
Note that in the adiabatic case, $q=0$, we have $\gameff=\gamma$, so we
recover
the standard stability criterion, $\gamma>4/3$.
In the radiative case, $\gameff \neq \gamma$, we finally obtain the
generalized stability criterion: 
\be
\label{eq:stability_criterion}
\gameff >\frac{2\gamma}{\gamma+{2}/{3}}\equiv \gamthr
\ee
For a monoatomic gas, where the adiabatic value is $\gamma=5/3$, 
the threshold for stability is $\gamthr=10/7=1.43$,
which is close but not identical to the adiabatic threshold $4/3$.

\subsection{Stability in terms of pre-shock quantities}

Next, we wish to express $\gameff$ and the stability criterion in terms of 
the properties of the pre-shock gas; the infall velocity $u_0$ and the
gas density $\rho_0$ at $\rs$. We use the jump conditions, \equ{j1} through 
\equ{j4}, in \equ{gameff}.  In \equ{j2} we assume $\us=0$, namely that 
the shock is temporarily at {\it rest},
which should be valid when the shock is marginally stable (or unstable).
This is because a stable shock is pushed outwards by the post-shock gas,
while cooling reduces the pressure, slows the outward motion,
and eventually causes it to halt and then be swept inwards
by the infalling matter and gravitational pull. 
The transition from stability to instability can thus be 
associated with a transition from expansion to contraction
of the shocked volume. 

According to \equ{ideal_eos} and \equ{j3} we have 
\be
\label{eq:e1}
e_1=\frac{1}{(\gamma-1)}\frac{P_1}{\rho_1}=\frac{2
u_0^2}{(\gamma+1)^2}\,.
\ee
According to \equ{drodt} and \equ{j2} with $\us=0$ we have 
\be
\frac{\dot\rho}{\rho}=-\frac{3u_1}{\rs}
=-3\frac{(\gamma-1)}{(\gamma+1)} \frac{u_0}{\rs} \, .
\ee
With these and \equ{q_def} 
we obtain the desired expression for the effective $\gamma$ of 
the post-shock gas in terms of the pre-shock conditions:
\be
\label{eq:final_gamma}
\gameff =\gamma -\frac{(\gamma+1)^4}{6(\gamma-1)^2}
\frac{\rho_0 \rs \Lambda(T_1)}{|u_0|^3} \, .
\ee
For a monoatomic gas, $\gamma=5/3$, we obtain 
\be
\gameff=5/3 -18.96 \frac{\rho_0 \rs \Lambda(T_1)}{|u_0|^3} \, .
\label{eq:gamma_mono}
\ee
Based on \equ{stability_criterion},
the criterion for stability of a $\gamma=5/3$ gas finally becomes
\be
\frac{\rho_0 \rs \Lambda(T_1)}{|u_0|^3} < 0.0126 \, .
\label{eq:fin_stability}
\ee

The post-shock temperature is related to the pre-shock infall velocity
using the jump condition, \equ{j4}, which for $\gamma=5/3$ gives 
\be
T_1 = \frac{3}{16} \frac{\mu}{k_bN_a} u_0^2 \, .
\label{eq:T1}
\ee
For a given cooling function $\Lambda(T)$,
\equ{fin_stability} is a simple criterion for determining whether a stable
shock can form at some radius $\rs$ of the halo. 
It is in a form that can be directly tested against hydrodynamic
simulations 
(\se{code}), and can serve for evaluating shock
stability under realistic conditions in cosmological haloes
(\se{cosmology}). 

\begin{figure}
\centerline{ \hbox{
\epsfig{file=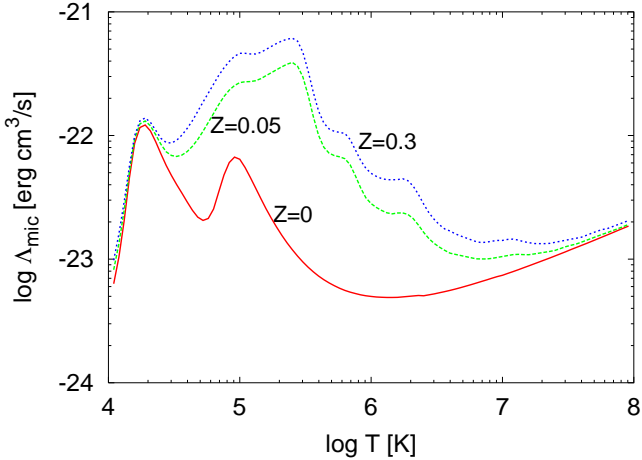,width=9cm}
 } }
\caption{The microscopic cooling function of low density 
gas for three different mean metallicities as indicated in solar units.
The peaks are dominated by atomic H and He (for $Z=0$) and by line
emission from heavier atoms (for the higher $Z$ values)}
\label{fig:lambda}
\end{figure}

Under the simplifying assumption that the gas is unclumped, the cooling
rate is given by \equ{q_def}. The macroscopic cooling function
$\Lambda(T)$ is related to the microscopic $\Lambda_{\rm mic}(T)$,
the energy-loss rate of a particle,
via $\Lambda(T) = (N_a^2 \chi^2/\mu^2)\Lambda_{\rm mic}(T)$,
where $\chi$ is the number of electrons per particle.
We assume a Helium atomic fraction of 0.1 for $\mu$ and $\chi$. 
The microscopic cooling function is shown in \Fig{lambda} for 
three different values of mean metallicity $Z$.
The cooling at temperatures below $10^4$K is very slow
because the main available cooling agent is molecular hydrogen, 
which is very inefficient.
At temperatures slightly above $10^4$K the cooling function peaks
due to \lya emission from atomic hydrogen.  
At very low metallicities, a second peak arises near $10^5$K due to 
recombination of atomic Helium.
Metals give rise to a higher peak at $\sim 10^5$K and slightly above,
due to line emission from the heavier atoms. 
At $\sim 10^6$K and above, the cooling is dominated by brehmstralung, 
and the cooling function increases slowly.
We use the cooling function as derived by \citet{sutherland:93},
and presented in their table, in the manner described in \citet{somerville:99}.

\section{The spherical hydro code}
\label{sec:code}

We test the validity of the shock stability criterion 
using numerical simulations based on a spherical hydrodynamics code
which follows the evolution of shells of dark matter and gas. 
Since the problem we intend to examine is of global spherical symmetry,
and since we need to follow the cooling and the shock with high precision,
we use a one-dimensional code.
Most of the simulations presented here were run using
$2000$ gas shells and $10,000$ dark-matter shells. 
A comparable resolution in a three-dimensional code
would require on the order of $10^{10}$ and $10^{12}$ 
particles respectively, which is impractical. 
We use no smoothing in the dark-matter shell-crossing scheme,
we introduce small-scale smoothing at the 
halo centre to avoid an artificial singularity there,
and we include small artificial viscosity in the hydrodynamics.
Tests of the code performance are described in Appendix \ref{app:code}.
 
\subsection{Dark matter}

The dark-matter particles are represented by infinitely thin spherical
shells of constant mass and of radii $r$ that vary in time. 
The shell of current radius $r$ obeys the equation of motion
\be
\label{eq:darkmatter}
\frac{d^2r}{dt^2}=-\frac{G(\md+\mg)}{(r+a)^2}+\frac{j^2}{r^3} \, ,
\ee
where $\md$ and $\mg$ refer respectively to the mass of dark matter and gas
within the sphere of radius $r$.
The last term is a centrifugal acceleration, determined by the
the specific angular momentum $j$ of the particle represented by the
shell. This $j$ is assigned to each shell at the initial conditions
and is assumed to be preserved during the simulation.
The parameter $a$ is the smoothing length that becomes effective 
only near the centre; it has been set to be $50$pc throughout this work.
The dark-matter shells are allowed to cross each other (and the gas
shells).
The dark-matter mass is evaluated by 
\be
\md(r)=\sum_{r_i <r} \Delta\md_i +
\frac{1}{2} \sum_j \delta(r=r_j) \Delta\md_j \, ,
\ee
where the shell radii and masses are denoted by $r_i$ and $\Delta\md_i$
respectively, $i=1, ...,\, n_{\rm d}$. 
The second term adds half the mass of a shell when $r$ 
coincides with one of the shells.
Generally, this summation requires $n_{\rm d}^2$ calculations for the dark
matter alone. The particles are kept sorted by radius. 
When two shells cross each other, we re-sort the
array by exchanging pairs which violate the order. 
This kind of sorting algorithm, termed
`Shell's Method' in Numerical Recipes \citep{NR97}, 
is natural in cases where only a few shells cross each other 
in each timestep. 
When two shells cross, they
exchange an energy of $G\, \Delta\md_i\, \Delta\md_j /r$. 
In order to conserve energy, the radius at which the shells cross 
must be known with great precision. We therefore reduce
the timestep to a small value, $t_{\rm sc}$, when two shells are about to cross
each other (see below).

\subsection{Gas}

The hydrodynamic part of the code is based on Lagrangian finite elements
in the form of spherical shells.
The basic equations governing the dynamics of each shell are
\be
\label{eq:force}
\frac{d^2r}{dt^2}=-\frac{1}{\rho}\nabla
(P+\sigma)-\frac{G(\md+\mg)}{(r+a)^2}+\frac{j^2}{r^3} \,,
\ee
\be
\label{eq:energy}
\frac{de}{dt}=\rho^2(P+\sigma)\frac{d\rho}{dt}-q \, ,
\ee
\be
\label{eq:density}
\rho=\frac{d\mg}{4\pi r^2 dr} \, . 
\ee
\be
P=(\gamma-1)\,e\,\rho  \, ,
\ee

An artificial viscosity term, $\sigma$, is added to the pressure for
numerical purposes, as explained below.
The smoothing length effective at the center, $a$, 
is the same as for the dark matter, \equ{darkmatter}.
As in the model described before,
the loss of internal energy due to radiative cooling is represented
by the cooling rate $q$.

The gas is divided into discrete shells. 
The mass enclosed within a shell, $\Delta\mg$, is assumed constant in time,
while the inner and outer shell boundaries move independently in time.
Each boundary is characterized by a temporal position $r$,
velocity $v$ and specific angular momentum $j$. 
The acceleration [\equ{force}] is evaluated at the boundary position.
The variables $\rho$, $P$, $e$, $q$, and $T$ for each shell
are evaluated within the shell between the boundaries.  

In particular, the pressure term in \equ{force} is evaluated at the 
outer boundary $r_i$ of shell $i$ using \equ{density}:
\be
-\frac{1}{\rho}\nabla(P+\sigma)=
\frac{4\pi r_i^2}{\Delta\mg} [(P+\sigma)_{i+1}-(P+\sigma)_i] \, ,
\ee
where $\Delta\mg=(\Delta\mg_i+\Delta\mg_{i+1})/2$.

The boundary conditions for the outer boundary of the system
are $P=\sigma=0$, and zero mass beyond the outer boundary.

Since gas shells cannot cross each other, the gas mass in the sphere interior
to each gas shell is constant throughout the simulation: 
\be
\mg(r_i)=\sum_{j=1}^{i}\Delta\mg_j \, . 
\ee
For the evolution of the dark-matter shell at $r$, 
we evaluate the gas mass that appears in \equ{darkmatter} using
\be
\mg(r)=\sum_{j=1}^{i-1}\Delta\mg_j
+\frac{r^3-r_{i-1}^3}{r_i^3-r_{i-1}^3}\Delta\mg_i \, ,
\ee
where $i$ refers to the gas shell for which $r_{i-1} \leq r <r_i$.

\subsection{Integration and Timestep}
\label{sec:integration}

The discrete integration of $r$ and $v$ is performed by a Runge Kutta
fourth-order scheme \citep{NR97}.
The state of the system at the beginning of each timestep is kept in memory
until the timestep is completed, such that it is possible to return to the 
beginning of the timestep and retry with a smaller timestep if the 
convergence criteria are not met. The timesteps are set such that
the position $r$ 
and velocity $v$ do not change by too much during a single timestep.
For a given accuracy parameter $\epsilon_{\rm rk}$, we demand that
the difference between the forth-order displacement $\Delta r_4$   
and the analogous first-order displacement $\Delta r_1$ obeys
$|\Delta r_4- \Delta r_1|/r < \epsilon_{\rm rk}$, 
both for the dark-matter and the gas.
The similar requirement is applied to the change in velocity over a timestep.
If this condition is not fulfilled, we reduce the timestep by a certain factor
and repeat the calculation over this timestep.
We use here as our default $\epsilon_{\rm rk}=0.1$.

In addition, we make sure the timestep for each shell does not violate 
the Courant condition, for an accuracy parameter $\epsilon_{\rm c}$. 
This implies 
${c_{\rm s}\Delta t}/{\Delta r} < \epsilon_{\rm c}$, where
$c_{\rm s}^2=(dP/d\rho)_s=\gamma P/\rho$
is the speed of sound.
We use here as our default $\epsilon_{\rm c}=0.3$.

A third limitation on the timestep comes from the desire to
conserve energy when shells cross. 
When two shells are about to cross each other within the current timestep $dt$,
we set the timestep to $\min (dt,t_{\rm sc})$, and keep it small until they 
actually cross. 
We use here as our default $t_{sc}=10^{-4}$Gyr.

The values for $\epsilon_{\rm c}, \epsilon_{\rm rk}$ and $t_{sc}$ were
chosen empirically such that energy is conserved and the dynamics 
converges to our satisfaction, in the sense that it does not change 
by much when smaller parameters are used.
We demonstrate in Appendix \ref{app:code} how well these requirements
are met.

Once we have computed the new radii and velocities of the shells 
at the end of the timestep, we correct the energy of the gas 
for the $-PdV$ work term using the states of the system at the beginning
and at the end of the timestep. 
The cooling is explicitly subtracted from the internal energy
after the hydrodynamic timestep is completed. Once the final
state of the system is ready, it is copied onto the memory array of the
initial state, and the simulation is ready to execute a new timestep.

\subsection{Initial conditions}
\label{sec:init}

The simulation starts at high redshift, $z=100$,
with a small spherical density perturbation. 
The initial density fluctuation profile is set to be
proportional to the linear correlation function of the assumed
cosmological model, representing the typical perturbation under the
assumption that the random fluctuation field is Gaussian 
\citep[see][and Appendix \ref{app:initial_perturbation}]{dekel:81}.
The amplitude of the density fluctuation at the initial time,
averaged over a given mass, determines the time of collapse, as desired.
The initial velocity field is assumed to follow a quiet Hubble flow and
the radial peculiar velocities build up in time.
We assume the standard $\Lambda$CDM cosmology with $\Omega_{\rm m}=0.3$,
$\Omega_\Lambda=0.7$, $h=0.7$ and $\sigma_8=1$. 

\subsection{Angular momentum}
\label{sec:centri}

We assume that in a real system the orbits of dark-matter particles, 
and the initial orbits of the gas particles, are quite elongated. 
Cosmological N-body simulations show that the velocity distribution 
tends to be more radial than tangential \citep{ghigna:98,safran:03}
and already for an isotropic distribution the eccentricities are about 1:6.
The processes we study in this paper occur away from the galactic disc
at a radius on the order of the virial radius, namely in a regime where 
the centrifugal force can be expected to be negligible compared to the 
gravitational force and the gas pressure force.
The prescribed angular momentum for the shells is thus mainly for 
numerical purposes, to avoid divergent densities of gas or dark matter 
shells when they pass through the halo centre.
Our results concerning the virial shock are insensitive to the actual
way by which we assign angular momentum to each shells.

In the current study we practically assume that the dark-matter
particles are almost on radial orbits.
The angular momentum of the gas is prescribed such that the shells, once they
lost their energy by radiation, would settle into an exponential disk 
with pure circular motions and a characteristic radius of a few kpc, 
smaller than the inner characteristic radius of the halo.
Our spherical `disc' thus contains gas that is cold and dense 
compared to the shocked gas. 

\subsection{Artificial viscosity}
\label{sec:visc}

It is impossible to follow the
discontinuous behavior across the shock using the conventional 
continuity equation for the density and standard conservation of
energy and momentum. The jump conditions can be calculated
explicitly, as in \equ{j1} to (\ref{eq:j4}) (termed `the
characteristic method' or `Godonov's method'). 
Alternatively, as proposed by Von-Newman, one can slightly smear 
the discontinuities and then solve them within the framework of the standard
hydrodynamic equations. 
By adding an artificial pressure term in a few shells around the shock, the
differential equations become solvable and 
one can continue the calculation without affecting the energy and the
dynamics of the shock (while its internal structure naturally changes). 

Artificial viscosity is applied when the inner and outer shell boundaries 
at $r_1$ and $r_2$ approach each other, $\Delta v =v_2-v_1 <0$,
and when the volume of the shell decreases,  
${dV}/{dt}=4\pi(r_2^2v_2-r_1^2v_1) <0$.
The artificial viscosity then takes the form
\be
\label{eq:av}
\sigma=a_2\rho(\triangle v)^2+a_1\rho c_s|\Delta v| \,.
\ee
The quadratic, common form of artificial viscosity smears
discontinuities over about 3 shells. The Linear discontinuity affects a
slightly larger range, and is usually added with a smaller coefficient $a_1$.
The coefficients $a_1$ and $a_2$ are varied for different shells in the course
of the simulation in order to overcome a specific numerical problem in the 
cold `disc', where the gravitational and centrifugal forces balance each
other and the pressure force is negligible. In this case the
gas is not a standard hydrodynamic gas because the
pressure does not regulate large discontinuities, and information is not
transported because of the low sound speed.
When a `disc' shell vibrates, it is artificially heated by the artificial
viscosity in every contraction until its pressure grows and stops the process. 
If we are not careful to properly tune the artificial 
viscosity we may end up with one `disc' shell that has been heated to $10^7$K
while the rest of the `disc' is at $10^4$K. This imposes an undesired
drastic decrease in the corresponding timestep.
In order to overcome this numerical problem, we gradually
turn off the quadratic term ($a_2$) of the artificial viscosity inside the
`disc'. 
We define a `disc' radius $\Rd$ to be the largest radius for which 
the difference between the gravitational and centrifugal forces is less
than 1/4 of the gravitational force.
Once at $r<0.6\Rd$, we continuously decrease the parameter $a_2$ in
\equ{av} according to $a_2=(r-0.3\Rd)/(0.3\Rd)$ and make it completely 
vanish at $r<0.3\Rd$. This prescription was found by trial-and-error to 
properly solve the numerical problem in most cases.
The linear term of the viscosity, being proportional to
the speed of sound, is anyway very small in the cold `disc', 
so effectively no artificial viscosity is applied in the inner `disc'.

Appendix \ref{app:code} provides tests and examples of the hydrodynamic 
simulations in some detail.

\section{Virial shock in the simulations}
\label{sec:simu}

\subsection{Existence of a virial shock}
\label{sec:simu1}

We now investigate the formation of a virial shock using
the spherical hydrodynamical simulations described above.
We wish to test in particular the validity of the analytic stability 
criterion developed in \se{analytical}.

\begin{figure}
\centerline{ \hbox{
\epsfig{file=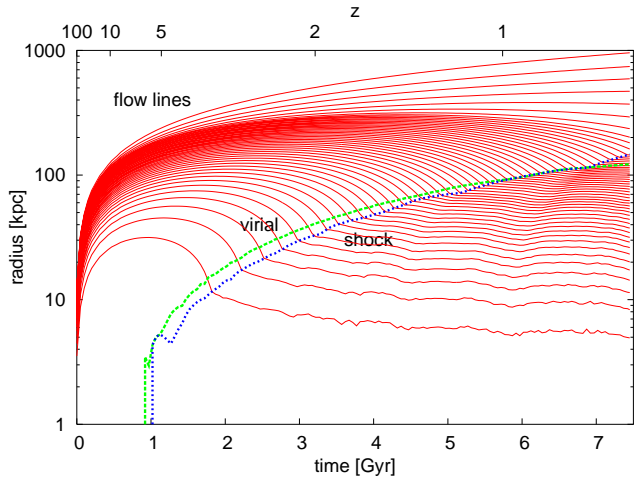,width=9cm} } }
\caption{Simulation of the {\it adiabatic} case. 
The sequence of (solid red) curves describe the log radii of Lagrangian 
gas shells as a function of time. The simulation was of $2000$ gaseous shells
(shown here) and $10,000$ dark matter shells. The radius of every 20th 
gaseous shell is plotted. 
Shown on top are the virial radius and the shock.
The shock exists at all times. It gradually propagates outwards, and it
practically coincides with the virial radius.}
\label{fig:rt_nocool}
\end{figure}

In order to mimic a typical perturbation in a random Gaussian field
(Dekel 1981, Appendix \ref{app:top_hat}),
the initial density-fluctuation profile was set to be proportional to the
correlation function, normalized such that the mean density fluctuation
in a sphere enclosing $M_{\rm i}=10^{11}\solm$ was 
$\delta_{\rm i}=0.09$ at $z=100$. 
For example, the shell encompassing $M \sim 3\times 10^{10}\solm$ 
is expected to 
collapse at $z=3$, and $M \sim 10^{12}\solm$ is expected to collapse at $z=0$.

\Fig{rt_nocool} shows the time evolution of the radii of Lagrangian
shells in a simulation of the adiabatic case, with the cooling turned off.
We find that a shock exists at all times. 
It appears as a sharp break in the flow lines, associated with a 
discontinuous decrease in infall velocity [\equ{j2}].
Shown in the figure is the shock radius, defined by the outermost shell for
which the inner and outer shell boundaries approach each other
and the volume of the shell decreases [the same conditions that have
been used for turning on the artificial viscosity in \equ{av}].
The shock gradually propagates outwards, encompassing more gas mass and dark
matter in time.
The gas below the shock is pressure supported and at quasi-static
equilibrium.
Not shown here are the dark-matter shells, which collapse, oscillate and
tend to increase the gravitational attraction exerted on the gas shells.

Shown in comparison is the evolution of the virial radius,
computed from the simulation density as the radius within 
which the mean overdensity is $\Dvir$ times the mean cosmological
background density.  The virial overdensity $\Dvir$ is provided by the 
dissipationless spherical top-hat collapse model;
it is a function of the cosmological model, and it may vary with time.
For the Einstein-deSitter cosmology, the familiar value is $\Dvir\simeq 176$
at all times.\footnote{This can be derived from the top-hat formalism 
of Appendix \ref{app:top_hat}, once the final radius is assumed to be
fixed at half the maximum-expansion radius but the overdensity is evaluated
at the time when the top-hat sphere would have collapsed to a singularity.}
For the family of flat cosmologies ($\omm+\oml=1$),
the value of $\Dvir$ can be approximated by \citet{bryan:98}
\be
\Dvir \simeq (18\pi^2 + 82x - 39x^2)/(1+x) \, ,
\label{eq:dvir}
\ee
where $x\equiv \omm(z)-1$, and $\omm(z)$ is the
ratio of mean matter density to critical density at redshift $z$.
For example, in the \lcdm\ cosmological model that serves as the basis for
our analysis in this paper ($\omm = 0.3$, $\oml=0.7$), the value at $z=0$ is
$\Dvir\simeq 340$.
We see in \fig{rt_nocool} that the shock radius almost coincides
with the virial radius at all times. 
This is hardly surprising, as the shock is likely to appear at the 
outermost radius at which shell crossing first occurs, which is
near the virial radius (to be demonstrated in \Fig{traj} below).

\Fig{rt_cool} is the result of a similar simulation, but now with
realistic radiative cooling for $Z=0$. 
We see that a stable shock does not exist in this case before $t=3.9$Gyr.
During this period, the cooling makes the gas lose its pressure support
and lets it collapse freely under gravity into the halo centre.
The collapse stops by the assumed angular momentum, in a `disc' whose
marked radius can be identified at the bottom of the plot
by the abrupt change of the infalling flow lines into horizontal lines.
The matter in the 'disc' is angular-momentum supported. As is visible
in the figure, a shock, in the sense of a discontinuity in velocity
and density, is present at the edge of the disc.
Once the stability criterion is met, a shock forms and propagates outward
abruptly. The propagation of the shock causes it to re-enter a regime for which
$\gameff$ is below the critical value. Consequently, the post shock gas becomes
non-supportive again and falls. The oscillatory behavior of the shock 
continues with an increasing period until it stabilizes at the largest 
radius for which the stability criterion is met. 
The shock never expands beyond the virial radius because shells
do not tend to cross there (\Fig{traj} below). 

\begin{figure}
\centerline{ \hbox{
\epsfig{file=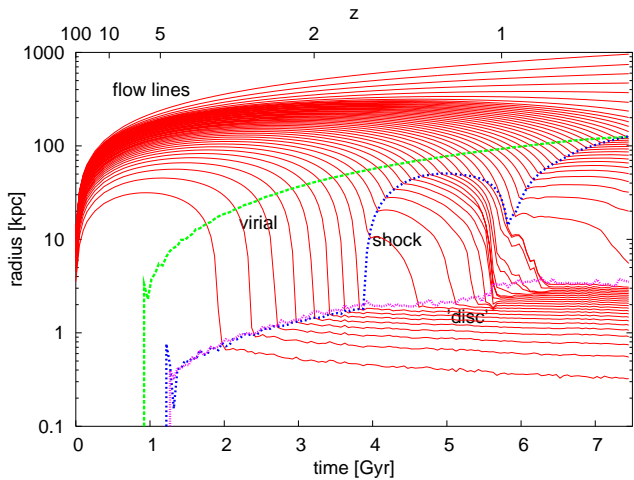,width=9cm} } }
\caption{
Simulation of the {\it radiative cooling} case, with $Z=0$.
The curves are as in \Fig{rt_nocool}, with the `disc' radius added.
There is no shock outside the `disc' at early times, 
when the virial mass is small, because the cooling is too efficient. 
A shock develops at later times, when the mass is larger,
and it quickly propagates outwards.
After a couple of oscillations the shock radius approaches the virial
radius.
}
\label{fig:rt_cool}
\end{figure}

\begin{figure}
\centerline{ \hbox{
\epsfig{file=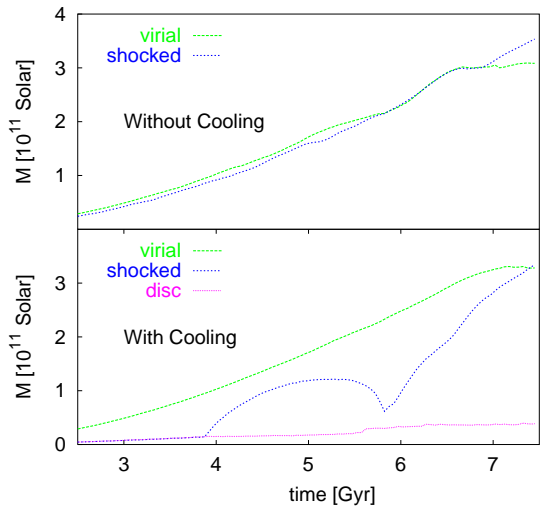,width=7.5cm} } }
\caption{Evolution of the total mass interior to the
virial radius, shock radius and `disc' radius, in the adiabatic simulation 
(top) and the radiative simulation (bottom).
}
\label{fig:2mass}
\end{figure}

\Fig{2mass} shows the evolution of the total mass interior to
the characteristic radii in the adiabatic
and radiative simulations of \fig{rt_nocool} and \fig{rt_cool}.
In the adiabatic case, the shock mass practically coincides
with the virial mass, and the gas never forms a disc.
In the radiative case, the shock is initially at
the disc radius, and only after 3.9Gyr does it start propagating outwards. 

\subsection{Testing the stability criterion}
\label{sec:comparison}

We first test the assumption made in \se{cool} that the post-shock gas has 
a homologous velocity profile in the vicinity of the shock. In Figures
\ref{fig:rt_nocool} and \ref{fig:rt_cool} we see that the flow lines beneath 
the shock tend to be nearly parallel in the $\log r - t$ plane. This means 
that $d\log r/dt=u/r \simeq $const., namely homology, \equ{homo}. 
A similar behavior has been found in all the simulations that we have 
performed. 

\begin{figure}
\centerline{ \hbox{
\epsfig{file=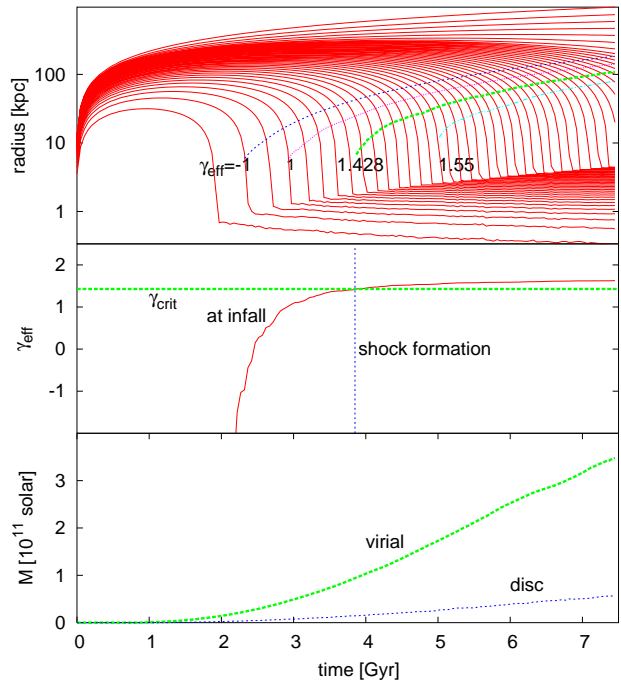,width=9cm} } }
\caption{Model predictions for shock stability.
{\it Top:} 
The flow lines describe a simulation similar to that shown in
\fig{rt_cool}, except that the cooling rate has been set to be 
unrealistically high such that the shock cannot develop.
The `disc' radius is marked by the break in the flow lines at the bottom.
Shown on top are 4 contours of equal $\gameff$ values (dashed lines,
for $\gameff=-1.0,\, 1.0,\, 1.428,\, 1.55$)
They are computed by the model, [\equ{final_gamma}], from the local 
(``pre-shock'') gas quantities corresponding to the flow lines at the
background, but assuming realistic cooling in the computation of $\gameff$.
The long-dashed (green) curve corresponds to  
$\gameff=\gamthr\simeq 1.428$, above which a shock would have formed
under realistic cooling.  
The negative value, $\gameff=-1$, is not physical.
{\it Middle:}
Time evolution of $\gameff$ at infall just above the `disc' radius (solid)
in comparison with the model critical value for stability $\gamthr=1.428$.
Since $\gameff$ is monotonically increasing with decreasing radius,
the shock is expected to form first at the `disc' radius.
Shock formation is predicted by the model at $t\simeq3.9$Gyr. 
{\it Bottom:}
Evolution of the characteristic masses in the simulation with the
unrealistically high cooling rate. 
The virial mass at the time of shock formation is predicted
to be about $10^{11}\solm.$}
\label{fig:supercool}
\end{figure}

Next, we use the simulations to test the validity of the stability criterion 
derived in \se{analytical}, \equ{fin_stability}, 
where $\gameff$ is given by \equ{final_gamma}.
Recall that the $\gameff$ is expressed in terms of the pre-shock 
quantities.  In order to map the value of $\gameff$ in the different 
regions of the free falling gas in the forming halo, we ran the same 
simulation as in the previous section except that the cooling rate was set 
to be very high, such that the virial shock never develops.
\Fig{supercool}, top panel
 shows the flow lines in this case, 
on which overlaid are 4 contours of equal $\gameff$ values, evaluated via
\equ{final_gamma} with \equ{T1} and the cooling rate from \citet{sutherland:93}.
As shells are falling into the halo, their $\gameff$ is gradually increasing.
Also, as time progresses, the value of $\gameff$ at the same radius is
increasing. By following the value of $\gameff$ just above
the ``disc'' radius (shown as the break at the bottom of the plot),
in comparison with the critical value for stability $\gamthr$,
we can therefore use our model to predict when we expect the virial shock
to form.  This is shown in the middle panel of
\fig{supercool}.
We see that at early times (and smaller masses) we have $\gameff<\gamthr$, 
predicting no stable shock.  The system is predicted to enter the stable-shock
regime at about $t=3.9$Gyr, where $\gameff$ becomes larger than $\gamthr$.
A comparison with the realistic radiative simulation described in
\fig{rt_cool} yields that this model prediction is very accurate: the shock 
indeed starts forming at $t=3.9$Gyr, and is globally stable thereafter.

\begin{figure}
\centerline{ \hbox{
\epsfig{file=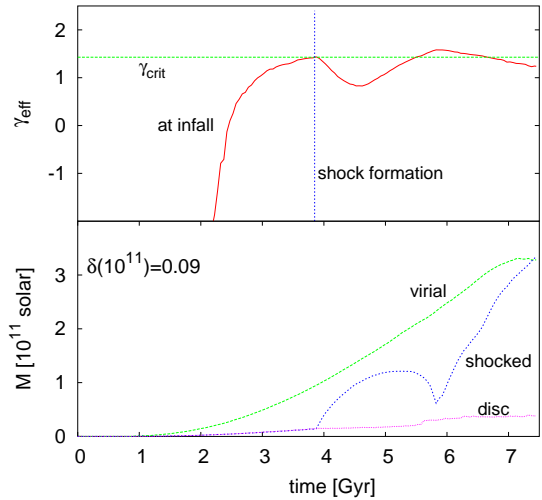,width=7.5cm} } }
\caption{Time evolution of $\gameff$ at 
the `disc' radius or the shock radius, whichever is larger (top), 
and the associated
characteristic masses (bottom), in the simulation with realistic cooling
of \fig{rt_cool}.
The shock forms for the first time when $\gameff$ becomes larger
than $\gamthr=10/7$. The shock then oscillates while it approaches
a steady state near the virial radius.}
\label{fig:halo1}
\end{figure}

\Fig{halo1} shows the actual evolution of $\gameff$ 
at either the `disc' radius or the shock radius, whichever is larger, 
as computed directly from the pre-shock quantities
in the simulation with realistic cooling. 
Shown at the same times are the
characteristic masses, already shown in \fig{2mass}.
The virial shock is first generated at time $t=3.9$Gyr, 
when the virial mass is $10^{11}\solm$. 
Starting at this time, the shock is propagating outwards very rapidly.
As a result of this fast expansion, the $\gameff$ of the pre-shock
infalling matter
at the shock, which is decreasing with $r$, drops below the threshold. 
This makes the shock lose its pressure support, it becomes
temporarily unstable and its expansion slows down until it is eventually
swept back on a dynamic time scale.  The associated drop in total mass behind the shock, seen around $t=5.8$Gyr,
is due to the fact that the dark matter is not swept back with the gas.
Once the shock is shrunk to a low enough radius, $\gameff$ rises again to above
$\gamthr$; the shock becomes stable again and it resumes its associated
expansion towards the virial radius. After the conditions for the
shock stability are first met, the shock is visible most of the
time. In the rest of this paper, we treat haloes at this state as ones
containing a virial shock. 

\begin{figure}
\centerline{ \hbox{
\epsfig{file=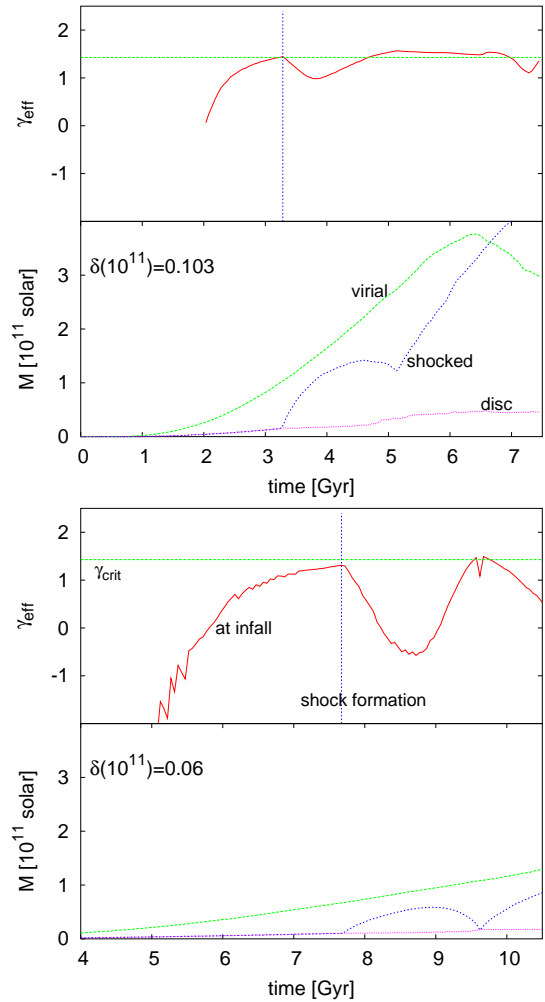,width=7.5cm} } }
\caption{Same as \fig{halo1} for two other initial conditions.
In all 3 cases the cosmology is \lcdm and the initial fluctuation density
profile is proportional to the correlation function, but with different
amplitudes, as indicated at $z=100$
within the sphere enclosing $10^{11}\solm$.
Note the different time scale in the bottom panel.
The qualitative behavior of the shock is similar in the three cases,
and so is the success of the model in predicting shock stability.
}
\label{fig:8mass}
\end{figure}

\Fig{8mass} presents results similar to \fig{halo1} for two other
simulations with different initial overdensities, and therefore 
different masses collapsing at different times.
The small difference seen in one case between the $\gameff$ at which 
the shock actually forms and the predicted $\gamthr$ may be totally due to
numerical inaccuracies in the simulation. Such inaccuracies may occur
when a dark-matter shell crosses a gaseous shell, which,  
near the threshold, may lead to a slightly premature shock formation.
This is seen in the convergence test of our code described
in Appendix~\ref{app:code}, when we compare the shock formation times 
in table \ref{tab:divs}.  Thus, the model stability criterion is found to 
be valid within the accuracy of the simulations 
in all the cases studied, indicating that
the model is not limited to a special range of masses and collapse times.

\section{Shock stability in cosmology}
\label{sec:cosmology}

The analysis of \se{analytical} thus provides a successful
criterion for shock stability, 
\equ{fin_stability}, as a function of the pre-shock properties of the
infalling gas at radius $r$: the density, velocity and metallicity.
In order to apply this criterion to a given protogalaxy in a cosmological
background, we wish to evaluate the gas density and velocity just before
it hits the disc, for a gas shell initially encompassing a total mass $M$
that virializes at redshift $\zv$.
In this calculation
we assume an Einstein-deSitter cosmology, as a sensible approximation at 
$z>2$ (where $\omm > 0.9$).

We assume a given universal baryonic fraction $\fb$, 
and a global spin parameter $\lambda$ which determines the ratio of disc to 
virial radius. 
The initial mean density perturbation profile, $\bar\deli(M)$,
is given at some fiducial time in the linear regime;
it is the average profile derived from the power spectrum of initial 
density fluctuations, 
as described in Appendix \ref{app:initial_perturbation}.
In the cosmological toy model used here we approximate the
power spectrum as a power law, $P_k \prop k^n$, where 
$n \simeq -2.4$ to mimic the $\Lambda$CDM power spectrum on galactic scales.

We follow gas shells from the initial perturbation till they approach the disc
using a two-stage model.
During the expansion, turn-around, and until an assumed {\it virialization} at
half the maximum-expansion radius, we assume no shell crossing, 
the total mass interior to the shell remains constant in
time, and we follow the radii, density and velocity of the shell via the
spherical top-hat model (see Appendix \ref{app:top_hat}). 
From the virial radius inwards we assume that the gas shells, 
which do not cross
each other, contract inside the fixed potential well of an isothermal 
dark-matter halo. 
This idealized model involves several crude approximations,
such as the instantaneous transition at the virial radius, and neglecting the 
effect of the angular momenta of the individual gas particles at small radii, 
but we show using spherical simulations that 
this model predicts the minimum halo mass for which a stable shock
first appears to an accuracy better than 25\%. 
This allows us to use the model for exploring the critical mass
as a function of cosmological parameters such as galaxy formation time,
metallicity, spin parameter, fluctuation power spectrum, and baryonic fraction.

\subsection{Toy model until virialization}
\label{sec:cos_1}

For a given shell $M$ and initial mean perturbation profile $\delta(M)$
[standing for the $\bar\deli(M)$ of Appendix \ref{app:top_hat}],
the top-hat model [\equ{r_m} and \equ{t_eta}] yields the implicit solution
\be
r(M,\eta)= \rv(M)\, (1-\cos \eta) \, ,
\label{eq:r_m_eta}
\ee
\be
t(M,\eta)= \tv(M)\, (\eta -\sin\eta) \, ,
\label{eq:t_m_eta}
\ee
where the mass dependence enters via the virial quantities 
\be
\rv=\ci\, M^{1/3} \delta(M)^{-1} \, , 
\ee
\be
\tv= \rv/\vv= \ci^{3/2} G^{-1/2}\, \delta(M)^{-3/2} \, ,
\ee
with $\vv^2=GM/\rv$.
The coefficient $\ci=(6/\pi)^{1/3} (0.15/\rhui)$ is determined by
$\rhui$, the cosmological density at the initial time when $\delta(M)$ 
is given, independent of $M$.

The velocity of the shell $M$ is 
\be
u = \frac{\partial r/\partial \eta}{\partial t/\partial \eta}
= \vv \frac{\sin\eta}{(1-\cos\eta)} \, .
\ee
At virialization, $\eta=3\pi/2$, it is simply $u=-\vv$.

In order to evaluate the local density, we
follow the radii of two adjacent shells, encompassing masses $M$ and $M+dM$
respectively, at a given time $t$, e.g., the time when shell $M$ virializes 
(at half its maximum expansion radius).  Let $\eta$ correspond to shell $M$ 
at that time, and $\eta+d\eta$ to shell $M+dM$. 
In order to express $d\eta$ in terms of $dM$ we use the fact that the time $t$
is the same for the two shells:
$0=dt=(\partial t/\partial M)dM +(\partial t/\partial \eta) d\eta$.
Using \equ{t_m_eta} this gives
\be
d\eta = \frac{3}{2} \frac{(\eta-\sin \eta)}{(1-\cos\eta)}
\frac{\delta'}{\delta} dM \, ,
\label{eq:deta}
\ee
where we denote $\delta' \equiv d\delta/dM$.
Expressing $dr$ in terms of $dM$ and $d\eta$ based on \equ{r_m_eta},
we obtain using \equ{deta} and after some algebra
\be
\frac{dr}{r} = \frac{1}{3}\frac{dM}{M} 
\left( 1- \frac{3M\delta'}{\delta} 
\left[ 1- \frac{3}{2}\frac{\sin\eta(\eta-\sin\eta)}{(1-\cos\eta)^2} \right]
\right) \, .
\label{eq:dr_dm}
\ee
At virialization of shell $M$, $\eta=3\pi/2$, the quantity in square brackets 
equals $(10+9\pi)/4$.
Not surprisingly, if the initial perturbation is of uniform density,
$\delta'=0$, we are left with $dr/r=(1/3)(dM/M)$, the straightforward result
of $M\prop r^3$.
Recall that the virial radius $\rv$ of shell $M$ can be obtained either 
from the universal density at the time of virialization using \equ{rho_eta},
or from the initial perturbation using \equ{r_m}.
The desired local density $\rho$ can be obtained from \equ{dr_dm}
via $dM = 4\pi r^2 \rho dr$.

If the initial perturbation profile is a power law, 
$\delta(r) \prop r^{-(n+3)}$,
using $M \prop r^3$ we have $3M\delta'/\delta=-(n+3)$.
So finally
\be
\frac{dr}{\rv} = \frac{1}{3}\frac{dM}{M}
\left( 1 + (n+3) \frac{(10+9\pi)}{4} \right) \, .
\label{eq:drv_dm_n}
\ee

\equ{dr_dm} [or \equ{drv_dm_n} in the power-law case]
allows us to compute the desired radii of the two adjacent shells at the
time of virialization of shell $M$.

\subsection{Toy model after virialization}
\label{sec:cos_2}

\begin{figure}
\centerline{ \hbox{
\epsfig{file=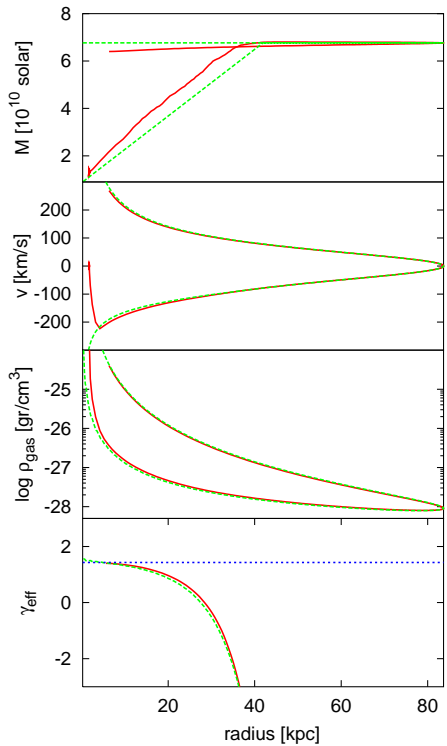,width=6.0cm} } }
\caption{Model versus simulation.
Evolution of the pre-shock gas properties along a trajectory of
one Lagrangian shell as a function of the radius. The solid (red) curve
corresponds to the smallest shell from the simulation of \fig{rt_cool} 
for which the shock formed. 
The dashed (green) curve is the prediction of the
toy model, calibrated here in an ideal way 
to match the simulation at maximum expansion.
The bottom panel shows the $\gameff$ derived from the above quantities
for the model and the simulation, in comparison with the
critical value of $\gameff$ marked by the horizontal line.
We notice that indeed $\gameff=\gamthr$ as $r \rightarrow 0$.
The fit between model and simulation is remarkable.}
\label{fig:traj}
\end{figure}

\begin{figure}
\centerline{ \hbox{
\epsfig{file=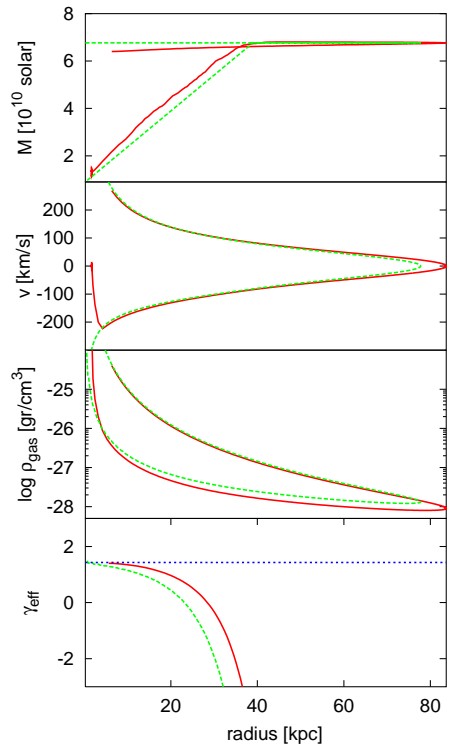,width=6.0cm} } }
\caption{Model versus simulation.
Same as \fig{traj}, except that the model was
calibrated in a practical way
based on $M$ and $\zv$ without reference to the
simulation results.
The model predictions now deviate somewhat from the simulation results,
but the deviation amounts to less than 25\% in the estimate of 
the critical mass.}
\label{fig:traj2}
\end{figure}

Given the radii of the two adjacent shells at $\rv$, we enter the
shell-crossing regime and continue to follow the shells down to the
disc radius by numerical integration.
The shell radius $r$ and velocity $u$ are related via energy conservation.
Assuming that the gas shells contract without crossing each other
inside a dark-matter halo that is a fixed isothermal sphere,
the total mass interior to the shell that originally encompassed a total
mass $M$ is 
\be 
M(r)=\fb M+(1-\fb)\frac{M}{\rv}r \, ,
\ee 
and the gravitational potential at $r$ is
\be 
\phi(r)=-\frac{GM(r)}{r}
-(1-\fb)\frac{GM}{\rv}\ln\left(\frac{\rv}{r}\right) \, .
\ee
 
The integration is performed by advancing $r$
according to the velocity $u$ and then recalculating $u$ according to 
energy conservation:
\be
(1/2)u^2+\phi(r)=(1/2)\vv^2+\phi(\rv)={\rm const.} 
\ee
We follow shell $M$ for the time it falls from $r=\rv$ to the disc radius
$r=\lambda \rv$,
and shell $M+dM$ for the same time interval. Denoting the separation between 
the shells at the end of this time interval by $dr$, we compute the
desired gas density by 
\be
\rho=\frac{\fb}{4\pi r^2} \frac{dM}{dr} \, . 
\ee
The resultant values of $r$, $u$ and $\rho$ are inserted into 
\equ{gamma_mono} in order to obtain an approximation for $\gameff$ 
and then to evaluate stability by \equ{stability_criterion}.
This allows us to check stability for the case where mass $M$ virializes
at redshift $\zv$, with metallicity $Z$, spin parameter $\lambda$, baryonic
fraction $\fb$, 
and a given power spectrum.

\subsection{Model versus simulations}

In \fig{traj} and \fig{traj2} we compare the evolution of the quantities
of a given gas shell according to the toy model described in the previous 
subsections and according to the spherical hydro simulation described in 
the earlier sections. We follow a specific shell
that hits the disc at about $t\simeq 3.8$Gyr, just before the shock
starts propagating into the halo [see \fig{rt_cool}].
The quantities shown as a function of radius $r$
are total mass $M$ interior to $r$,
radial velocity $u$, gas density $\rho$, 
and the corresponding value of $\gameff$.
For $M$ and $u$ the evolution starts at the top-left corner and ends at the
bottom left, while for $\rho$ the upper part of the curve corresponds to 
the expansion phase and the lower part to the contraction phase.
The evolution of $\gamma$ is followed only during part of the contraction
phase.
  
In \fig{traj} we calibrate the toy model to match the simulation at 
the maximum expansion radius. We see that while the mass interior to the
shell is reproduced by the model only to a limited accuracy in the last
stages of the collapse, the velocity, density and the resulting
value of $\gameff$ are recovered very well by the model.
This allows us to predict quite accurately the point where $\gameff$
exceeds $\gamthr$.

Since we wish to use the toy model without an exact knowledge of the
conditions at maximum expansion, we normalize the model evolution in 
\fig{traj2} based on $M$ and $\zv$.

The slight deviations in $u$ and in $\rho$ now translate into a larger
error in $\gameff$.  The error in the toy model originates mostly
from the slight ambiguity in the definition of the virial radius.
On one hand we assume it to equal half the maximum-expansion radius,
and on the other hand we assume it to represent an overdensity of $\Dvir$
as in \equ{dvir}. These two assumptions are not fully consistent
with the actual behavior of the virializing system in the simulation. 
Nevertheless, we see below that our approximate model allows us
to estimate the critical halo mass below which the shock does not form
to an accuracy of better than 25\%, which is quite satisfactory for our
purpose here.

\subsection{Critical mass for shock formation}

\begin{figure}
\centerline{ \hbox{
\epsfig{file=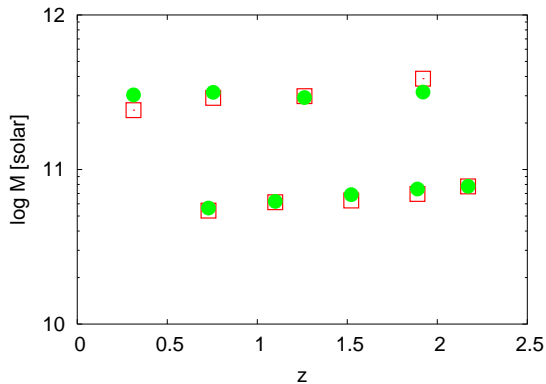,width=7.5cm} } }
\caption{Model (green circles) versus simulations (red squares).
Displayed is the critical halo mass for shock formation versus 
the virialization redshift.
In all cases $\fb=0.13$, $n=-2.4$, and $0.02<\lambda<0.05$. 
The high set of points is for $Z=0.05$ and the low set is for $Z=0$.
}
\label{fig:run_comp}
\end{figure}

\Fig{run_comp} shows for several different cases the critical halo mass,
below which a shock does not propagate into the halo, versus the redshift 
at which this critical mass virializes.
For each case we compare the model prediction
to the shock formation as actually seen in the simulation.
The cases differ by the mean metallicity, $Z=0$ and $Z=0.05$ for the lower and
upper sets of points respectively, 
and by the amplitude of the initial perturbation, corresponding to a range
of shock-formation redshifts at every given $Z$.
The assumed baryonic fraction is always $\fb=0.13$, but 
the assumed spin parameter may be different for the different shells in a given
simulation because we set it for each shell such that the final disc 
has an exponential surface density profile. However, the $\lambda$ values
vary in the range $0.02$ to $0.05$, compatible with the distribution of
spin parameter in cosmological simulations (Bullock \etal 2001).
We see that the model predicts the critical mass with an accuracy better than
25\%, such that we can use it for mapping the parameter space in more detail.

\begin{figure}
\centerline{ \hbox{
\epsfig{file=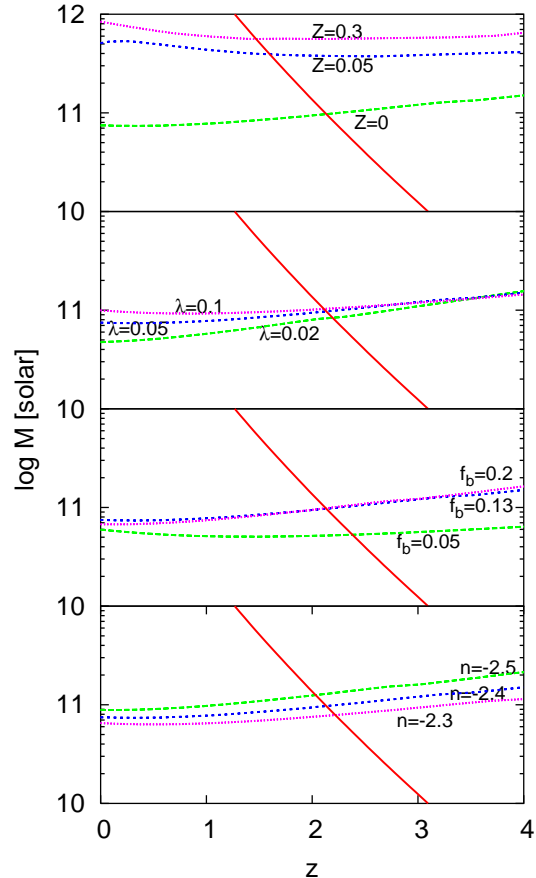,width=7.5cm} } }
\caption{Critical mass for shock formation as a
function of virialization redshift, according to the model tested against
simulations in \fig{run_comp} (dashed curves). 
Shown for comparison is the characteristic press-Schechter $M_\star$ 
as a function of $z$ for $\Lambda$CDM (solid curve).
The default values of the parameters are:
$\fb=0.13$, $\lambda=0.05$, $Z=0$, and $n=-2.4$.
One parameter is varied in each panel as indicated.
}
\label{fig:mz}
\end{figure}

\Fig{mz} shows for several different choices of parameters
the critical halo mass for shock formation versus the halo
virialization redshift as predicted by the model. 
A virial shock does not form in haloes of masses below the line.
The lines are not always monotonic due to the non-monotonic features
in the cooling curves (\fig{lambda}).
Shown in comparison is $M_\star$, the characteristic mass for haloes 
forming at $z$ according to the Press-Schechter approximation 
\citep{lacey:93}.  
The default values of the parameters, used unless specified otherwise,
are $Z=0$, $\lambda=0.05$, $\fb=0.13$ and $n=-2.4$. 

The upper panel has the metallicity varying from $Z=0$ to $Z=0.3$.
The critical mass tends to be higher at higher redshifts (especially for
$Z=0$) because the higher density implies more efficient cooling.
It is striking that even for the case of zero metallicity, for which the
cooling is not at its maximum efficiency, an $M_\star$ halo cannot
produce a shock until a relative late redshift, $z \simeq 2.1$. 
The addition of a small amount of metals, $Z=0.05$, increases the cooling 
rate significantly (see \Fig{lambda}) such that $M_\star$ haloes start 
producing virial shocks only after $z \simeq 1.6$.

The second panel has $\lambda$ varying as marked. 
The shock forms slightly earlier if the
disc is smaller (lower $\lambda$), because the 
conditions become more favorable for shock formation closer to the centre. 
At high redshifts the increase in infall velocity happens to
balance out the increase in density, temperature and cooling rate. 
The post shock temperature there is a few $10^6$K. 
The bottom line is that the critical mass is not too sensitive to $\lambda$. 

The third panel has $\fb$ varying as marked.
The critical mass is monotonic with the baryonic
fraction because the cooling rate is monotonic with gas density.
The parameter $\fb$ can be interpreted as the fraction of the baryons
that actually take part in the shock formation. This can be smaller than
the universal baryonic fraction if some of the gas falls into the halo
in the form of dense clumps.
Even with $\fb$ as low as $0.05$, meaning that most of the gas is not
participating in the cooling, an $M_\star$ halo would not produce a shock
until $z \simeq 2.4$.
The conclusion is that the critical mass is not too sensitive to $\fb$ either.

The bottom panel explores three values for the initial power index $n$ 
approximating the power spectrum of $\Lambda$CDM on galactic scales.  
The dependence of the critical mass on $n$ in this regime is weak. 

\section{Discussion}
\label{sec:conc}

The heating of the gas behind a virial shock in haloes has been a basic
component in galaxy formation theory \citep{rees:77}.
We studied the conditions for the existence of such a virial shock 
in spherical haloes. We first pursued an analytic stability analysis  
in the presence of cooling, and then demonstrated its validity
using high-resolution spherical hydrodynamical simulations.
The obtained criterion for shock stability in terms of the post-shock
quantities is 
\be
\gameff \equiv \frac{d\ln P/dt}{d\ln \rho/dt} > 10/7 \, .
\ee
In terms of the pre-shock gas properties, this condition reads
\be
\frac{\rho_0 \rs \Lambda(T_1)}{|u_0|^3} < 0.0126 \, ,
\ee
where $\rho_0$ and $u_0$ are the gas density and infall velocity 
in front of the shock, $\rs$ is the shock radius, $\Lambda(T)$ is the
cooling function which depends on the metallicity $Z$, 
and $T_1 \prop u_0^2$ is the post-shock temperature as a function
of the pre-shock infall velocity.

Based on this criterion,
we find that a virial shock forms only in big haloes forming at late 
redshifts.  A 
virial shock does not form in smaller haloes forming early where 
the cooling behind the shock efficiently removes its pressure support.
For example, we find that most galactic haloes that have collapsed and
virialized by $z\sim 2$ did not produce a virial shock. 
Haloes less massive than $\sim 10^{11}\solm$ never produce a shock
even if the gas is of zero metallicity. If the metallicity is non-negligible
(e.g. $Z\sim 0.05$), this lower bound to shock formation rises to
$\sim 7 \cdot 10^{11}\solm$.
When a shock does not exist, the gas 
is not heated to the halo virial temperature until it falls all the
way to the disc at the inner halo.

\citet{forcada:97}, in an unpublished work, have pursued independently
a numerical analysis along similar lines, involving a more  
detailed treatment of the cooling processes involved.
They also find that the virial shock radius is significantly reduced due to
the cooling in haloes of small masses, $M<10^{11}\solm$. 
In their case the shock never completely disappears because of a different
feature in their numerical scheme; they put all the cooled post-shock gas
in one central ``shell" to avoid numerical difficulties at the centre.
This makes the inner boundary of the system follow the shock quite closely
in cases where there is efficient cooling behind the shock, and allows
the presence of a small-radius shock even in such cases.
Overall, our numerical results are in encouraging agreement,
and our analytic model provides a natural explanation for their numerical
results as well.

The most severe uncertainty when attempting to apply our results to
real galaxies arises from the assumed spherical symmetry
in both the model and the numerical simulations. 
The validity of this approximation
for the asymmetric halo configurations in the hierarchical clustering scenario
is an open question to be addressed in future work.
Nevertheless, we notice that \citet{katz01} and \citet{fardal:01} 
observe in their
cosmological simulations that a large fraction of the mass accreted onto
haloes indeed remains cold and is never heated to the virial temperature.
\citet{toft:02} find in their simulations, using a similar treeSPH
code as the one 
used by \citet{katz01}, that the soft X-ray radiation is mainly emitted from 
within the innermost $20kpc$ of their haloes, well inside the virial radius,
in encouraging agreement with our results.
On the other hand, it is not obvious that the resolution in these
simulations is adequate for studying the shock physics involved;
our estimates indicate that three-dimensional simulations with proper 
resolution are not practical at present (Appendix \ref{app:code}).

Another complication may arise from radiative effects. 
Even when there is no virial shock,
the kinetic energy of the gas eventually turns into radiation
when the gas infall motion is brought to a halt at the disc. 
At such densities, the width of the shock front is much smaller than 
the width of the cooling front behind the shock, $\sim 10^{-2}$pc versus 
$\sim 10^2$pc.
Thus, the gas in a thin shell behind the shock is heated to a temperature 
corresponding to its kinetic energy, and it cools by radiating soft X-rays.
The X-ray radiation is expected to generate an ionized $H_{\it II}$ bubble,
in which the ionization rate balances the recombination rate. 
The Str\"omgren radius of this bubble is relatively small, on the order of a 
few kiloparsecs, because the high gas density implies a high recombination 
rate.  The recombination process then generates a flux of \lya radiation, 
emitted at the inner few kpc of the halo.

A naive inspection of cross sections might indicate that the \lya radiation 
would be trapped inside the halo. This could in principle affect the shock
stability in three different ways: by increasing the radiation pressure, 
by heating up the infalling matter, and by slowing down the radiative 
cooling responsible for the shock instability.
It has been argued by \citet{rees:77} that the radiation pressure 
at these low temperatures must be insignificant compared
to the gas pressure even if all the internal energy was drained from 
the baryons into the radiation field.  
One might add that since the radiation pressure behaves like a $\gamma=4/3$ 
gas, it could at most make the system marginally stable.
When work is done on the radiation field, any leakage of
radiation out would turn the energy into cooling rather
than $PdV$, and will thus reduce the effective gamma,
making the system unstable.

Partial heating of the infalling gas should not affect our analysis
as long as the temperature of the infalling matter is significantly below
the virial (post-shock) temperature such that the strong shock approximation
remains valid.
The effect of the reduced cooling rate is yet to be investigated.
In practice, we do not expect the radiation trapping to be very efficient, 
because the effective opacity is reduced by thermal broadening
and by the systematic blue shift due to the gas infall motion.
When the opacity is high, the radiation heats up the gas,
which enhances the thermal broadening and the collisional ionization rate.
This reduces the opacity and allows for radiation escape.
The system is likely to reach a steady state in which it gradually cools.
This process is under current investigation.

Feedback effects may further complicate the picture and affect shock
stability. The energy fed back to the gas from stars, supernovae
and AGNs may heat the halo gas and expel part of it. Merging
substructures may have additional complicated effects. These effects
cannot be captured by our idealized spherical analysis, and
a proper study would probably require high-resolution three-dimensional 
hydrodynamical simulations. 
While observations and certain theoretical considerations
indicate that feedback effects are likely to
be important in galaxies as large as $\sim 10^{11}\msun$ 
and may thus affect the shock stability (e.g. Dekel \& Silk
1986; Dekel \& Woo 2003, and references therein),
it has proven difficult for numerical simulations to reproduce such effects
in any but small dwarf galaxies, indicating that feedback effects
may not be so crucial for the understanding of shock stability.
Until the dust settles on the role of feedback effects,
our preliminary conclusions based on the spherical analysis
should be taken with a grain of salt.
%

The general absence of a virial shock 
might have 
three direct implications, which we study in associated papers.

First, as explained above,
when the gas is heated at the disc rather than near the halo
virial radius, the generated X-ray radiation serves to ionize the gas
and is not emitted outwards.
The result would be a suppression of the X-ray emission in the range 
$5\times 10^5$ to $2\times 10^6$K.
This may help explaining the missing 
X-ray problem pointed out by \citet{pen:99} and
\citet{benson:00}. \citet{pen:99} argue that there is 
an order-of-magnitude discrepancy between the soft X-ray flux as observed 
by \citet{cui:96}, after subtracting the contribution of quasars, 
and the predicted flux from haloes constructed by a Press-Schechter 
hierarchical model under the assumption of shock heating to the virial 
temperature. 

Second, the infall energy, via the ionizing X-ray, 
is efficiently transformed into \lya radiation at the inner few kpc of
the halo. 
A related increase in the \lya flux has indeed been seen in the cosmological
simulations of \citet{fardal:01}.
This may explain the observed high flux of \lya emitters at high redshift
\citep[e.g.][]{pentericci:00,pentericci:01,breuck:00}. 
Based on the high observed flux and the assumption that the \lya is emitted
from stars, \citet{pentericci:00} estimate large masses for the \lya emitters,
but the much higher flux per unit mass predicted by our model
may lead to significantly lower mass estimates.
Based on our analysis, most of the \lya flux is expected to be emitted
from the inner few kpc of the halo, where the gas is at $\sim 10^4$K.
Neglecting line shifts and broadening, the halo might be opaque to \lya,
thus eventually emitting its energy from an outer photosphere where 
the halo becomes transparent. However, a careful study of the thermal
broadening and the systematic redshifts within the halo is required
in order to determine whether the system is opaque or transparent to the
\lya photons. This is a subject of an ongoing investigation.

Finally,
the direct collapse of cold gas into the disc may have interesting theoretical
consequences to be worked out. It may induce an efficient star burst
in analogy to the burst originating in the shock between two
colliding gas clouds. In turn, the strong inwards flow of gas may
prevent an efficient gas removal by supernova-driven winds.
In particular,
current cosmological semi-analytic models (SAM) of galaxy formation
\citep[and related works]
{kauffmann:93,kauffmann:99,cole:94,somerville:99,maller:01} 
use the standard picture of heating behind a virial shock 
in their modeling. This has strong effects on the disk formation rate,
star formation rate, feedback etc. 
Other semi-analytic models \citep{efstathiou:00,white:91} 
also appeal to the slow gas infall rate as a mechanism that regulates
the gas input into the disc. Since the cooling time for a $10^{11}\solm$ 
halo is relatively short, the SAM predictions for such haloes may
be only slightly affected by the inhibition of heating.
However, given some metal enrichment, no heating is expected for haloes
as massive as $\sim 7\times 10^{11}\solm$, for which the cooling time is 
longer, and the effect on the SAM predictions may be more severe.
Shocks, when present, are also expected to alter the properties of the
gas, for example - extinct dust particles. These effects can change
SAMs that incorporate dust extinction.

\section*{Acknowledgments}
We acknowledge advice from Z. Barkat and E. Livne,
 J. Ostriker
and stimulating discussions with S. Balberg, 
E. Bertschinger,
T. Broadhurst, D. Gazit, Y. Hoffman, 
W. Mathews,    
A. Nusser, N. Shaviv, and S.D.M. White. 
This research has been supported
by the Israel Science Foundation grant 213/02,
by the German-Israel Science Foundation grant I-629-62.14/1999.
and by NASA ATP grant NAG5-8218.


\appendix
\section{Testing the hydro code}
\label{app:code}

The numerical code, {\it Hydra}, has been developed specifically 
for simulating the evolution of a single spherical halo through collapse 
and feedback processes. A proper computation of the cooling and shock
formation requires high precision.
In this appendix we describe a few of the tests performed in order to
verify that the code works properly. In the following three subsections
we test for energy conservation, spatial convergence, and the performance
of the code in a self-similar case.

\subsection{Energy Conservation}
\label{app:energy}
 
Our numerical scheme does not use the total energy equation in the integration
of the partial differential equations. 
Furthermore, the total energy of the system is not a straightforward
sum of other variables that are involved in the calculation.
The requirement of energy conservation is therefore an independent test
for the accuracy of the numerical scheme.
Energy conservation is harder to achieve than spatial 
convergence for several reasons. First, the error in total energy is 
systematic, in the sense that when dark-matter shells cross 
each other the energy tends to increase. Second, since  
our system is only marginally bound, the
total energy is a small difference between two large quantities. 
We notice that energy conservation is simpler to achieve when there is no 
cooling, or when dark matter is absent (and thus there is no shell crossing).

The total energy of the system at time $t$ is the sum of terms:
\be
E=K_{\rm d}+T_{\rm d} +K_{\rm g}+T_{\rm g}+U+Q \, ,
\ee
where subscripts d and g refer to dark matter and gas respectively,
$K$ stands for kinetic energy, $T$ stands for potential energy,
$U$ is the gas internal energy, and $Q$ is the thermal energy lost 
to radiation by time $t$.

For the dark matter, these are straightforward sums over the discrete
dark-matter shells:
\be
K_{\rm d}= \Delta \md_i
\frac{1}{2}\sum_{i=1}^{n_{\rm d}} \left(v_i^2+\frac{j_i^2}{r_i^2}\right)
\, ,
\ee
\be
T_{\rm d}=
 -\sum_{i=1}^{n_{\rm d}}\frac{G\, \Delta \md \, {\mt}_i } {r_i+a} \, ,
\ee
where ${\mt}_i$ is the total mass interior to dark-matter shell $i$,
as defined in \se{code}.

For the gas shells, recall that the quantities $r$, $v$ and $j$ are given at the
inner and outer shell boundaries, $i-1$ and $i$ respectively, so we
compute the shell energies by averaging over the two boundary values:
\begin{eqnarray}
K_{\rm g} =\!&\!&\! 
  \frac{1}{2} \sum_{i=1}^{n_{\rm g}} \Delta \mg_i
  \left(\frac{v_i r_i^3+v_{i-1}r_{i-1}^3} {r_i^3+r_{i-1}^3}\right)^2
\\ \!&+\!&\!
 \frac{1}{2} \sum_{i=1}^{n_{\rm g}} \Delta \mg_i
  \left(\frac{j_i r_i^2+j_{i-1}r_{i-1}^2} {r_i^3+r_{i-1}^3}\right)^2
 \, , \nonumber
\end{eqnarray}
\be
T_{\rm g}=
-\sum_{i=1}^{n_{\rm g}}
\frac{G\, \Delta \mg_i\, ({\mt}_i +{\mt}_{i-1})/2 }
     {[(r_i^3+r_{i-1}^3)/2]^{1/3} +a} \, .
\ee

The internal energy is a straightforward sum 
\be
U=\sum e_i \Delta \mg_i \, .
\ee
The energy radiated away, $Q=\int dt \int q dm$,
is computed by
\be
Q_=\sum_{j=1}^{n_{\rm t}} \Delta t^j 
\sum_{i=1}^{n_{\rm g}} q^j_i \Delta \mg_i \, ,
\ee
where $\Delta t^j$ is the length of timestep $j$,
and 
$q^j_i$ is the cooling rate in shell $i$ at timestep $j$ (in
units of ${\rm erg} \mbox{ }{\rm g}^{-1} {\rm s}^{-1}$).

In a run with 10,000 dark-matter shells and 2,000 gas shells, we require
and obtain energy conservation at the level of 1\% in a Hubble-time,
using a typical Runge Kutta timestep of about $5\times 10^{-6} \mbox{Gyr}$.
(Such a run takes about 10 hours on an Alpha-6 DEC processor).

We check the conservation first by varying the accuracy parameters presented
in \se{integration}, and then by varying the number of shells.
The three cases presented in Table \ref{tab:eps} demonstrate 
that the results converge when the accuracy is increased.
The simulations shown in this table are of the standard case with realistic
cooling shown in \fig{rt_cool}. 
When cooling is shut off, energy conservation is much better.
With the nominal choice of accuracy parameters the final energy is 
$0.9999$ of the initial energy.

\begin{table}
\caption{Energy conservation}
\begin{tabular}{@{}llllllc}
\hline
description & $n_{\rm g}$ & $n_{\rm d}$ & $\epsilon_{\rm c}$ & $\epsilon_{\rm
rk}$ &$t_{sc}/Gyr$ &
$E_{\rm fin}/E_{\rm init}$\\
\hline
nominal         & 2k & 10k & $0.3$ & $0.1$  & $10^{-4}$ & $0.992$ \\
small $\epsilon$'s & 2k & 10k & $0.1$ & $0.01$ & $10^{-5}$ & $0.994$ \\
more shells& 3k & 15k & $0.3$ & $0.1$  & $10^{-4}$ & $0.997$ \\
\hline
\end{tabular}
\label{tab:eps}
\end{table}
 
\subsection{Spatial Convergence: 3D versus 1D}

\begin{table}
\caption{Shock formation times and energy conservation}
\begin{tabular}{@{}llcc}
\hline
$n_{gas}$ & $n_{dark}$ & $E_{final}/E_{initial}$ & formation of shock(Gyr) \\
\hline
$3,000$ & $15,000$ & $0.997$ & $3.9$ \\
$2,000$ & $10,000$ & $0.994$ & $3.9$ \\
$1,000$ & $5,000$   & $0.976$ & $4.01$ \\
$500$  & $2,500$   & $0.923$ & $3.76$ \\
$250$  & $1,250$   & $0.711$ & $3.27$ \\
$125$  & $625$    & $0.252$ & $2.82$ \\
\hline 
\end{tabular}
\label{tab:divs}
\end{table}

A proper treatment of the competition between the pressure increase due to
contraction and the pressure decrease due to cooling requires high 
temporal 
and spatial resolution. In particular, when the 
spatial resolution is increased, the shock appears earlier.
Table \ref{tab:divs} shows results from simulations of the 
case with realistic cooling (\fig{rt_cool}), all with the same 
accuracy criteria ($\epsilon_{\rm c}$, $\epsilon_{\rm rk}$, and $t_{sc}$), 
but with different spatial resolutions. 
The average distance between gas shells near the center ranges from
about $80$pc to $2$kpc. With the poorest resolution of 125 gas shells
the virial shock appears almost immediately after the virialization of 
the first shells of the simulation. The energy changed by about 75\%
during this simulation.
Even if we assume that the precision of a 3D calculation is as good as that
of an analogous 1D calculation (actually SPH codes 
converge slower than finite element schemes for problems 
involving shocks), we still need to cube the
number of particles or grid points in order to achieve the same resolution. 
A three-dimensional simulation with 
$2\times 10^6$ gas particles and $2.5\times 10^8$ dark-matter particles, 
which is close to the limit of what is computationally feasible today,
would correspond to the unsatisfactory case with the lowest spatial 
resolution in table \ref{tab:divs}. 

\subsection{A self-similar case}

When the initial conditions are scale free (unlike the
initial conditions assumed in the body of this paper,
motivated by $\Lambda$CDM), and when the cooling function
is also scale free (unlike the realistic cooling function used above),
the results should be self similar. 
This can provide a test for the accuracy of
our numerical code.
We follow 
\citet{bertschinger:85a,bertschinger:85b} 
in using an initial perturbation 
consisting of a point-mass embedded in a uniform-density background.
Far from the point mass, the system should be self similar. 
We ran a simulation of such a case using our code with gas only ($\fb=1$)
and no cooling, starting at $z=200$ with an overdensity of 10 inside the 
innermost $2$kpc.

\begin{figure}
\centerline{ \hbox{
\epsfig{file=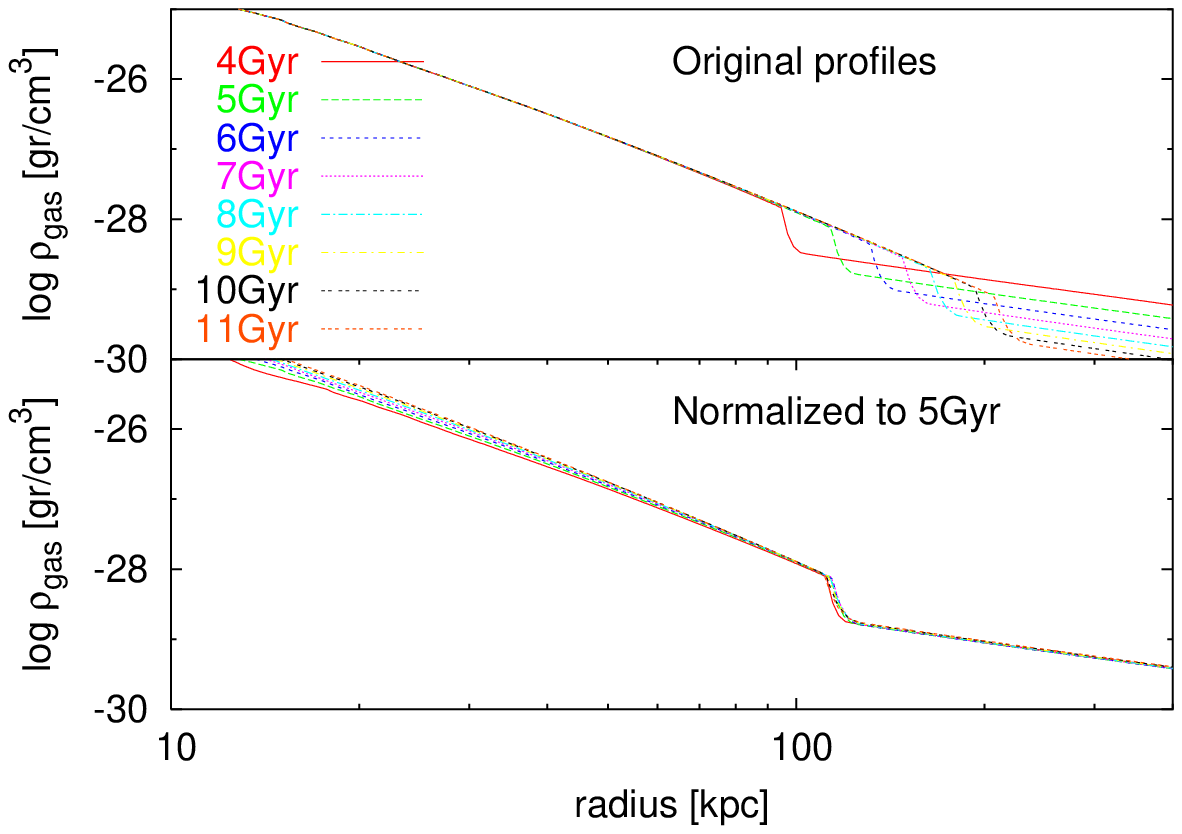,width=9cm}
} }
\caption{Self-similar profiles by our hydro code with gas only.
{\it Top:} Density profiles at times $4$ to $11$Gyr.
{\it Bottom:} The same profiles
normalized according to the self-similarity law $r\prop t^{8/9}$ to $t=5$Gyr.}
\label{fig:self}
\end{figure}

\begin{figure}
\centerline{ \hbox{
\epsfig{file=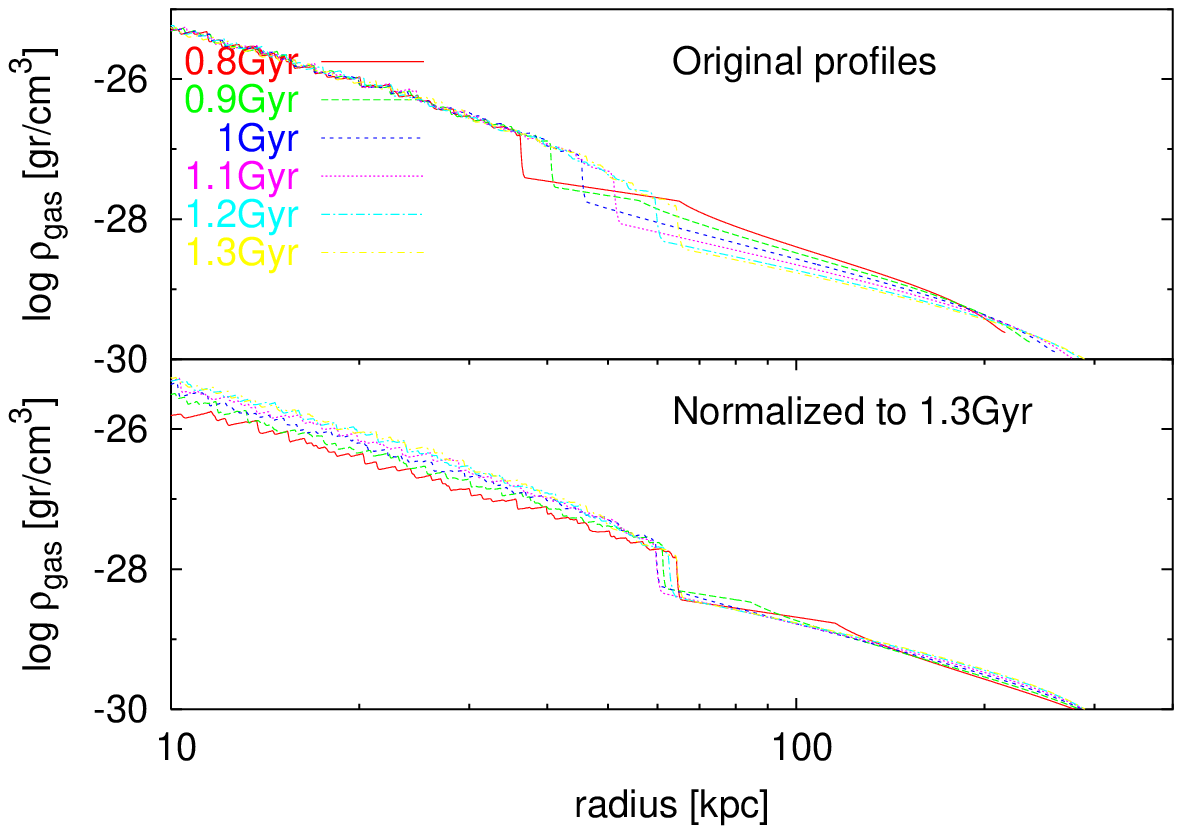,width=9cm}
} }
\caption{Similar to \Fig{self} but with dark matter included, $\fb=0.13$.
The profiles are scales to $t=1.3$Gyr.} 
\label{fig:self2}
\end{figure}

The upper panel of \fig{self} shows the density profile at different times.
As expected, a shock appears at every time as a density jump by a factor 
of 4 [$=(\gamma+1)/(\gamma-1)$ for $\gamma=5/3$], and 
the post-shock gas settles to a complete rest after it is shocked.
(The slope of the post-shock density profile is somewhat different from
\citet{bertschinger:85b}, because our calculation assumes a \lcdm cosmology
rather than the Einstein-deSitter assumed by Bertschinger.)
The lower panel shows the same profiles after they were
scaled to the same time (5Gyr) according to the scaling
relation of \citet{bertschinger:85b}: $r\prop t^{8/9}$.   
We see that our simulations recover the expected scaling relation
almost perfectly.

\Fig{self2} shows an analogous test for the case where both gas and dark 
matter are present, with $\fb=0.13$. The results are similar except for 
the somewhat higher noise 
level caused by the dark-matter component.

\section{Top-Hat Model}
\label{app:top_hat}

Consider a bound spherical perturbation encompassing mass $M$, 
whose mean density fluctuation profile at some fiducial
initial time in the linear regime is $\bar\deli(M)$, embedded
in an Einstein-deSitter (EdS) cosmological background when the universal
expansion factor is $\ai$.
We wish to express the shell radius $r$ as a function of time in terms
of $M$.

The implicit solution for a closed ``mini-universe", via a conformal time
parameter $\eta$ specific to this perturbation, is
\be 
r=\rv (1-\cos \eta) \, ,
\label{eq:r_eta}
\ee
\be
t=\frac{\rv}{\vv}(\eta-\sin \eta) \, .
\label{eq:t_eta}
\ee 
Maximum expansion occurs when $\eta=\pi$, and then a collapse to half this
radius, which we identify with {\it virialization}, is obtained at 
$\eta=3\pi/2$, 
with virial radius $\rv$ and corresponding virial velocity $\vv^2={GM}/{\rv}$.
We normalize the universal expansion factor $a$ by identifying it
at the initial time with the shell radius $r$. 
Assuming $\eti \ll 1$, we have 
$\ri\simeq (1/2)\rv \eti^2$ and $\ti\simeq(1/6)(\rv/\vv)\eti^3$,
so $\ai=\ri$ yields $\ai=(9\rv\vv^2/2)^{1/3}\ti^{2/3}$. 
The EdS expansion factor, $a \prop t^{2/3}$, can now be related 
using \equ{t_eta} to the perturbation's $\eta$ at any time:
\be
a=(9\rv\vv^2/2)^{1/3} t^{2/3} = (9/2)^{1/3}\rv (\eta-\sin \eta)^{2/3} \, .
\label{eq:a_eta}
\ee
The mean density within the perturbation relative to the universal density
at the same time becomes a straightforward function of $\eta$:
\be
\frac{\bar\rho}{\rhuu}=\frac{a^3}{r^3}
=\frac{9}{2} \frac{(\eta-\sin \eta)^2}{(1-\cos \eta)^3} \, .
\label{eq:rho_eta}
\ee
This is a standard result of the top-hat model.

In order to relate the density to the small initial perturbation
at $\eti\ll 1$, we obtain from \equ{rho_eta} by a proper Taylor expansion
to the first non-vanishing order:
\be
\bar\deli\simeq 0.15 \eti \, ,
\ee
where $\bar \delta \equiv \bar\rho/\rhuu -1$ is the mean fluctuation.
Using this in the linear term of \equ{r_eta} we obtain
\be
\ai=(1/2)\rv\eti^2 =(1/0.3)\rv \bar\deli \, .
\label{eq:a_i}
\ee
This allows us to write the mean density at any $\eta$, using \equ{r_eta}, as
\be
\frac{\bar\rho}{\rhui}=\frac{\ai^3}{r^3}
=\frac{\bar\deli^3}{0.3(1-\cos \eta)^3} \, .
\ee
Recalling that $\bar\rho = M/[(4\pi/3)r^3]$ we finally obtain at any $\eta$
\be
r= \ci \frac{M^{1/3}}{\bar\deli(M)} (1-\cos \eta) \, ,
\label{eq:r_m}
\ee
where 
\be
\ci \equiv \left(\frac{6}{\pi}\right)^{1/3} \frac{0.15}{\rhui} \, .
\ee
In particular, at $\eta=3\pi/2$, we obtain for the virial radius
$\rv = \ci {M^{1/3}}/{\bar\deli(M)}$.
The constant $\ci$ is independent of $M$; the universal density $\rhui$ 
[$=(1+\zi)^3\rhu0$] is determined by the choice of the fiducial
redshift $\zi$ at which $\bar\deli(M)$ is given.

\section{Initial Profile}
\label{app:initial_perturbation}

\def\del0i{\delta_{\rm 0i}}
We adopt in the linear regime the typical density fluctuation profile 
for the assumed power spectrum of fluctuations. For a Gaussian random
field, this profile is proportional to the two-point correlation function
\citep{dekel:81}:
\be
\deli(r)=\del0i \frac{\xi(r)}{\xi(0)} \, ,
\ee
where $\del0i$ specifies the amplitude normalization.
For a given power spectrum $P(k)$, the correlation function is given by
\citep[eq.~21.40]{peebles:93} 
\be
\xi(r)=4\pi \int_0^\infty k^2 dk P(k) \frac{\sin kr}{kr} \, ,
\ee
and the local variance is
\be
\xi(0)=4\pi \int_0^\infty k^2 dk P(k) \, .
\ee

The mean density fluctuation interior to radius $r$,
containing mass $M=(4\pi/3) \rhui r^3$ when the fluctuation is small, 
is 
\be
\bar\deli(r)=\frac{3}{r^3}\int_0^r r^2dr \deli(r) \, .
\label{eq:bardeli}
\ee
This involves the integral \citep[eq.~21.62]{peebles:93}
\be
J_3(r)\equiv \int_0^r r^2 dr \xi(r)
= \frac{4\pi r^3}{3} \int_0^\infty k^2 dk P(k) \tilde W_s(kr) \, ,
\ee
where $\tilde W_s(kr)$ is the Fourier transform of the top-hat window,
\be
\tilde W_s(kr) = 3\left[ \frac{\sin kr}{(kr)^3} - \frac{\cos kr}{(kr)^2}
\right]
=\frac{3}{kr}j_1(kr) \, ,
\ee
with $j_1$ the spherical Bessel function.

We use in the simulations of this paper the 
$\Lambda$CDM power spectrum as from the fitting formula of \citet{bardeen:86}
\be
P(k) = AkT^2(k),
\ee
\be
T(k)=\left\{1+\left[ak/\Gamma+(bk/\Gamma)^{3/2}+(ck/\Gamma)^2\right]^\nu \right\}^{1/\nu}
\ee
with $a=6.4h^{-1}$Mpc,  $b=3.0h^{-1}$Mpc, $c=1.7h^{-1}$Mpc, $\nu=1.13$ and $\Gamma=0.21$ \citep[the $\tau$CDM model of][]{efstathiou:92}.
The normalization is such that $\sigma_8=1.$

In the cosmological toy model we approximate the $\Lambda$CDM by 
a power-law power spectrum $P_k \prop k^n$,  for which
the two-point correlation function is also a power law, 
$\xi(r)\prop r^{-(n+3)}$, 
and then  
\be
\bar\deli(r)= \left( {r\over \r1i} \right) ^{-(n+3)} ,
\ee
where $\r1i$ provides the normalization.
In terms of mass we obtain
\be
\bar\deli(M) = \left( {M\over (4\pi/3) \rhui \r1i^3} \right) ^{-(n+3)/3} \, .
\label{eq:deli_m}
\ee
This serves as the input to \equ{r_m}, or \equ{dr_dm}.

We normalize the initial perturbation such that a specific mass 
$M_1$ reaches virialization at some cosmological epoch $\av=1/(1+\zv)$.
Using \equ{a_eta} and \equ{a_i} we obtain the linear analog to the nonlinear
fluctuation growth rate:
\be
\frac{\bar\delta(\eta)}{\bar\deli}
\equiv \frac{a(\eta)}{\ai}
=0.3 \left( \frac{9}{2}\right) ^{1/3} (\eta -\sin \eta)^{2/3} \, .
\ee
At virialization, this gives 
$\delv\equiv \bar\delta(\eta=3\pi/2) \simeq 1.58$.
Then:
\be
\bar\deli(M_1) = \delv \left( {\ai \over \av} \right) .
\ee
The normalization parameter $\del0i$ (or $\r1i$) at $\ai$ 
is obtained by equating this with \equ{bardeli} [or \equ{deli_m}] at $M=M_1$.

\label{lastpage}

\end{document}